%% file: main.tex
\documentclass[
   twocolumn, prd, aps,
   floatfix,
   nofootinbib,
]{revtex4-1}

\newcommand{\myTitle}{Analysis of the decay $D^0\rightarrow K_S^0 K^+ K^-$}

\input{config} 
\input{alias} 

\begin{document}

\title{\myTitle}
\date{\today}
\begin{abstract}
  \input{authors}
  Using a data sample of \SI{2.93}{\invfb} of \epem collisions collected at $\sqrt{s}=\SI{3.773}{\GeV}$ in the \bes experiment, we perform an analysis of the decay \DKsKK. The Dalitz plot is analyzed using \num{1856(45)} flavor-tagged signal decays. We find that the Dalitz plot is well described by a set of six resonances: $a_0(980)^0$, $a_0(980)^+$, $\phi(1020)$, $a_2(1320)^+$, $a_2(1320)^-$ and $a_0(1450)^-$. Their magnitudes, phases and fit fractions are determined as well as the coupling of $a_0(980)$ to \KKbar, $g_{\KKbar}=\SIe*{3.77}{0.24}[0.35]{\GeV}$.
  The branching fraction of the decay \DKsKK is measured using \nume{11660}{118} untagged signal decays to be \SIe*{4.51}{0.05}[0.16]{\timesten\tothe{-3}}. Both measurements are limited by their systematic uncertainties.
\end{abstract}
\maketitle 

\input{Introduction}
\input{DetectorData}

\input{Selection}
\input{AmplitudeAnalysis}
\input{BF}
\input{Conclusion}

\input{acknowledgements}
\FloatBarrier
\bibliography{inspire,additional}

\end{document}

%% file: config.tex
\usepackage{ifthen}
\newboolean{enable-lineno} 
\setboolean{enable-lineno}{false} 
\newboolean{enable-printout} 
\setboolean{enable-printout}{false} 

\usepackage{microtype} 
\usepackage[super]{nth} 
\usepackage{lmodern}
\renewcommand\rm{\textrm} 

\PassOptionsToPackage{pagewise}{lineno} 
\usepackage{lineno} 
\modulolinenumbers[1] 
\ifthenelse{\boolean{enable-lineno}}%
{%
   \newcommand*\patchAmsMathEnvironmentForLineno[1]{%
    \expandafter\let\csname old#1\expandafter\endcsname\csname #1\endcsname
    \expandafter\let\csname oldend#1\expandafter\endcsname\csname end#1\endcsname
    \renewenvironment{#1}%
    {\linenomath\csname old#1\endcsname}%
   {\csname oldend#1\endcsname\endlinenomath}}%
   \newcommand*\patchBothAmsMathEnvironmentsForLineno[1]{%
    \patchAmsMathEnvironmentForLineno{#1}%
   \patchAmsMathEnvironmentForLineno{#1*}}%
   \AtBeginDocument{%
    \patchBothAmsMathEnvironmentsForLineno{equation}%
    \patchBothAmsMathEnvironmentsForLineno{align}%
    \patchBothAmsMathEnvironmentsForLineno{flalign}%
    \patchBothAmsMathEnvironmentsForLineno{alignat}%
    \patchBothAmsMathEnvironmentsForLineno{gather}%
    \patchBothAmsMathEnvironmentsForLineno{multline}%
   }
   \linenumbers 
}{\nolinenumbers}
\usepackage{etoolbox}
\makeatletter
\patchcmd\linenumberpar{\@LN@parpgbrk}{\penalty\@LN@parpgpen\relax}{}{}
\makeatother

\usepackage{natbib}
\PassOptionsToPackage{fleqn}{amsmath}		
\usepackage{amsmath}
\usepackage{amsthm}
\usepackage{mathtools}

\usepackage{xfrac} 
\usepackage{nicefrac} 
\usepackage[makeroom]{cancel}
\usepackage{xparse} 
\usepackage{physics} 

\usepackage{siunitx} 
\usepackage{suffix} 
\usepackage{siunitx_extension} 

\usepackage{booktabs} 
\usepackage{adjustbox} 
\usepackage{multirow}
\usepackage{tabulary}
\usepackage{tabularx} 
\usepackage{enumitem} 
\PassOptionsToPackage{caption=false}{subfig} 
\usepackage{subfig}
\usepackage{floatpag}
\usepackage{rotating}
\usepackage{placeins} 

\usepackage{scrhack} 

\PassOptionsToPackage{novbox}{pdfsync} 
\usepackage{pdfsync} 
\PassOptionsToPackage{
   pdftex,
   hyperfootnotes=false,
   pdfpagelabels
}{hyperref}
\usepackage{hyperref}
\hypersetup{%
   pdfstartpage=3, pdfstartview=FitV,%
   colorlinks=true, linktocpage=true, 
   breaklinks=true, pdfpagemode=UseNone, pageanchor=true, pdfpagemode=UseOutlines,%
   plainpages=false, bookmarksnumbered, bookmarksopen=true, bookmarksopenlevel=1,%
   hypertexnames=true, pdfhighlight=/O,
   urlcolor=webbrown, linkcolor=RoyalBlue,
   citecolor=webgreen, 
   pdftitle={\myTitle},%
   pdfsubject={},%
   pdfkeywords={},%
   pdfcreator={pdfLaTeX},%
   pdfproducer={LaTeX with hyperref}%
}
\ifthenelse{\boolean{enable-printout}}%
{%
   \hypersetup{%
    colorlinks=false, linktocpage=false, pdfborder={0 0 0}, pdfstartpage=3, pdfstartview=FitV,%
   }
}{\relax}

\PassOptionsToPackage{pdftex}{graphicx}
\usepackage{graphicx}
\DeclareGraphicsExtensions{.pdf,.png,.gif,.jpg,.jpeg} 
\graphicspath{{figures/}}

\PassOptionsToPackage{english,capitalise}{cleveref}   
\usepackage{cleveref} 

\PassOptionsToPackage{T1}{fontenc} 
\usepackage{fontenc}
\PassOptionsToPackage{utf8}{inputenc}	
\usepackage{inputenc}
\usepackage{csquotes} 
\usepackage{textcomp} 
\usepackage{afterpage}

\PassOptionsToPackage{usenames,dvipsnames,svgnames,table}{xcolor} 
\usepackage{xcolor}
\definecolor{webgreen}{rgb}{0,.5,0} 
\definecolor{webbrown}{rgb}{.6,0,0} 
\definecolor{turquoise}{rgb}{0 206 209}

\usepackage{xspace} 
\usepackage{url} 
\usepackage{umoline} 

\usepackage{tikz}
\usetikzlibrary{calc, tikzmark, decorations.markings,
   custom, 
}
\usepackage{pgfplots}
\pgfplotsset{compat=newest} 
\usepackage{pgfplotstable}

\PassOptionsToPackage{
   colorinlistoftodos,
   textwidth=2.5cm,
   textsize=scriptsize
}{todonotes}
\usepackage{todonotes}

%% file: authors.tex
\begin{small}
\begin{center}
M.~Ablikim$^{1}$, M.~N.~Achasov$^{10,d}$, P.~Adlarson$^{60}$, S. ~Ahmed$^{15}$, M.~Albrecht$^{4}$, M.~Alekseev$^{59A,59C}$, A.~Amoroso$^{59A,59C}$, F.~F.~An$^{1}$, Q.~An$^{56,44}$, Y.~Bai$^{43}$
% ...
\\
\vspace{0.2cm}
(BESIII Collaboration)\\
\vspace{0.2cm} {\it
$^{1}$ Institute of High Energy Physics, Beijing 100049, People's Republic of China\\
% ...
$^{4}$ Bochum Ruhr-University, D-44780 Bochum, Germany\\
% ...
$^{10}$ G.I. Budker Institute of Nuclear Physics SB RAS (BINP), Novosibirsk 630090, Russia\\
% ...
$^{15}$ Helmholtz Institute Mainz, Johann-Joachim-Becher-Weg 45, D-55099 Mainz, Germany\\
% ...
$^{43}$ Southeast University, Nanjing 211100, People's Republic of China\\
$^{44}$ State Key Laboratory of Particle Detection and Electronics, Beijing 100049, Hefei 230026, People's Republic of China\\
% ...
$^{56}$ University of Science and Technology of China, Hefei 230026, People's Republic of China\\
% ...
$^{59}$ (A)University of Turin, I-10125, Turin, Italy; (B)University of Eastern Piedmont, I-15121, Alessandria, Italy; (C)INFN, I-10125, Turin, Italy\\
$^{60}$ Uppsala University, Box 516, SE-75120 Uppsala, Sweden\\
% ...
$^{d}$ Also at the Novosibirsk State University, Novosibirsk, 630090, Russia\\
% ...
}\end{center}

\vspace{0.4cm}
\end{small}

%% file: Introduction.tex
\section{Introduction}
\label{sec:intro}
The decay \DKsKK is a self-conjugate channel\footnote{Charge conjugation is implied throughout this work, except where explicitly noted otherwise.} with a resonant substructure containing \CP eigenstates as well as non-\CP eigenstates. 
An accurate measurement of the decay and its substructure has implications for various fields. 
The substructure of \DKsKK\footnote{Where beneficial we abbreviate the final state \KsKK by $3K$.} is dominated by the \KKbar S-wave which can be studied in an almost background-free environment. In particular, light scalar mesons are of interest since their spectrum is not free of doubt~\cite[p.~658ff.]{Tanabashi:2018oca}.
The branching fractions of the resonant substructure and the total branching fraction are inputs to a better theoretical understanding of \Dz -\Dzb mixing.
Furthermore, the strong phase difference between the decays \Dz\to\KsKK and \Dzb\to\KsKK can be determined from the amplitude model. This phase is an input to a measurement of the angle $\gamma$ of the CKM unitarity triangle using the decay of \Bm\to\Dz\Km with \Dz\to\KsKK~\cite{Aubert:2008bd}. A model-independent determination of the strong phase will be presented in a separate paper~\cite{BESIII-strong-phases-KsKK}.

The most recent analysis of \DKsKK was performed by the \babar experiment~\cite{Aubert:2005sm}. Using \num{12500} flavor-tagged \Dz decays, the total branching fraction was measured relative to the decay \Dz\to\KS\pip\pim. The current value given by the Particle Data Group (PDG)~\cite{Tanabashi:2018oca} is derived from that measurement. Furthermore, a Dalitz plot analysis was performed and it was found that the resonant substructure is well described by a set of four resonances: $a_0(980)^0$, $a_0(980)^+$, $\phi(1020)$ and $f_0(1370)$.

In this work, the decay \DKsKK is analyzed using a data sample of $\epem$ collisions corresponding to an integrated luminosity of \SI{2.93}{\invfb}~\cite{Ablikim:2014gna,*Ablikim:2015orh} collected with the \bes detector at $\sqrt{s}=\SI{3.773}{\GeV}$. At this energy, the produced \psiprpr decays predominantly to \DzDzb  and \DpDm. The pair of neutral \D mesons is produced in a quantum entangled state. The flavor of one meson can be inferred from the decay of the other meson if it is reconstructed in a flavor-specific decay channel. The subsample in which both \Dz's are fully reconstructed is referred to as the `tagged sample'. The sample in which only the reconstruction of the signal decay is required is denoted as the `untagged sample'.

The paper is structured as follows: we introduce the detector and the Monte-Carlo (MC) simulation in \cref{sec:detector} followed by the description of the event selection in \cref{sec:dataPrep}. The Dalitz plot analysis is presented in \cref{sec:dalitz} and the branching fraction measurement in \cref{sec:bf}. Finally, we summarize our results in \cref{sec:conclusion}.

%% file: DetectorData.tex
\section{Detector and data sets}
\label{sec:detector}
The \bes detector records symmetric \epem collisions with high luminosity\footnote{The current record is \SI{1.0}{\timesten\tothe{33}\centi\meter\tothe{-2}\second\tothe{-1}} at $\sqrt{s}=\SI{3.773}{\GeV}$} provided by the BEPCII storage ring~\cite{Yu:2016cof}. The center-of-mass energy ranges from \SIrange{2}{4.6}{\GeV}, and \bes has collected large samples in this energy region~\cite{Ablikim:2019hff}, in particular in the charmonium region above \SI{3}{\GeV}.
The detector covers \SI{93}{\percent} of the full solid angle. It is composed of the following main components: 
the helium-based multi-layer drift chamber (MDC) which is the most inner component of the detector provides momentum measurement of charged tracks as well as a measurement of the ionization energy loss \dedx. The momentum of a charged track with a transverse momentum of \SI{1}{\giga\eVc} is measured with a resolution of \SI{0.5}{\percent} and the \dedx resolution for electrons from Bhabha scattering is \SI{6}{\percent}.
A plastic scintillator time-of-flight (TOF) system provides a time resolution of \SI{68}{\ps} (\SI{110}{\ps}) in the barrel (end cap) part and is used for particle identification. 
The electromagnetic calorimeter (EMC) measures the energy of electromagnetic showers with a resolution of better than \SI{2.5}{\percent} and \SI{5}{\percent} at energies of \SI{1}{GeV} in the barrel and end cap parts, respectively. 
The outermost part is a system of resistive plate chambers for muon identification (MUC) interleaved in the iron return yoke of a superconducting solenoidal magnet that provides a magnetic field of \SI{1}{\tesla}.
A detailed description of the detector can be found in Ref.~\cite{Ablikim:2009aa}.

Monte Carlo (MC) simulated events are used to develop the selection criteria, estimate backgrounds, and to obtain reconstruction efficiencies. 
The simulation is based on \sftName{Geant4}~\cite{Agostinelli:2002hh} which includes the geometric description of the \bes detector and the detector response. The simulation includes the beam energy spread and initial state radiation (ISR) in the \epem annihilations.
The inclusive MC samples are simulated using KKMC~\cite{Jadach:2000ir} and consist of the production of \DDbar pairs, non-\DDbar decays of the \psiprpr, the ISR production of the \jpsi and $\Psi(3686)$ states, and continuum processes.
The known decay modes are modelled with \sftName{EvtGen}~\cite{Lange:2001uf,Ping:2008zz} using branching fractions taken from the PDG~\cite{Patrignani:2016xqp}, and the remaining unknown decays from the charmonium states with \sftName{LundCharm}~\cite{Chen:2000tv,Yang:2014vra}. The final state radiations (FSR) from charged final state particles are incorporated with the \sftName{Photos} package~\cite{RichterWas:1992qb}.
All samples correspond to a luminosity of 5-10 times that of the data sample. 
In addition, the Dalitz plot analysis requires a large sample of phase space distributed events of the signal channel. 
In this case, the decay \psiprpr\to\DzDzb with \Dzb\to\KsKK and \Dz\to(tag) is simulated. 
`Tag' refers to a number of flavor-specific channels that are used for flavor tagging (see \cref{tab:sel:tagYields}).
The measurement of the branching fraction requires an accurate decay model for the signal decay, and the result of the Dalitz plot analysis is used to generate appropriate signal events which are used to substitute signal events in the inclusive MC sample.

%% file: Selection.tex
\section{Data preparation}
\label{sec:dataPrep}
\subsection{Event selection}
\label{sec:dataPrep:sel}
Charged tracks are reconstructed from hits in the MDC and their momenta are determined from the track curvature.  We require that each track has a point-of-closest approach to the interaction point of \SI{10}{\centi\metre} along the beam line ($V_z$) and \SI{1}{\centi\metre} perpendicular to it ($V_r$). Furthermore, we require that the reconstructed polar angle is within the acceptance of the MDC of $|\cos\theta| < 0.93$. 
The particle species is determined from \dedx measured by the MDC and TOF information. 
The combined $\chisq(H)$ for a particle hypothesis $H$ is given by
\begin{align}
   \chisq(H) = \chisq_{\dedx}(H)+\chisq_{TOF}(H).
   \label{eqn:dataPrep:sel:pidChiSq}
\end{align}
Using the corresponding number of degrees of freedom, a probability $P_H$ is calculated, and we require that all kaon and pion candidates satisfy $P_\kaon> P_\pi$ and $P_\pi> P_\kaon$, respectively.

Two pions with opposite charge are combined to form a \KS candidate. The requirement on $V_r$ is removed and the requirement on $V_z$ is loosened to \SI{20}{\centi\metre} for these tracks. No particle identification requirement is imposed. The reconstructed \KS invariant mass is denoted by \mKS. The common vertex and the signed \KS candidate flight distance are estimated using a secondary vertex fit. The \chisq of the secondary vertex fit is required to be smaller than \num{100}. 
To suppress combinatorial background and decays of the type $\Kp\Km\pip\pim$, we require that \KS candidates have a ratio of flight distance over its uncertainty larger than \num{2} for the untagged sample and larger than \num{0} for the tagged sample. 
The flavor tag final states include \piz's and $\eta$'s which are reconstructed from their decays to a pair of photons~\cite{Ablikim:2018tay}.

We combine a \KS candidate and two oppositely charged kaon candidates to form a \Dz signal candidate. Due to the charm threshold decay kinematics, a convenient variable to discriminate signal from background is the beam-constrained mass
\begin{align}
   \mBC^2\si{\c\tothe{4}}=E_\text{beam}^2-\abs{\vec{p}_{D}}^2\si{\cSq}.
   \label{eqn:sel:mBC}
\end{align}
The combined 3-momenta of the daughter tracks in the rest frame of the \psiprpr is denoted by $\vec{p}_D$ and the beam energy by $E_\text{beam}$. A kinematic fit with the nominal \Dz mass as a constraint is applied, and candidates are required to have a \chisq smaller than \num{20}.
In the untagged sample, only the decay of one \Dz meson to \KsKK is reconstructed and the one with the smallest difference between the reconstructed energy of the candidate and half the center-of-mass energy of the \epem beams is chosen.
In the tagged sample, both \Dz and \Dzb decays are reconstructed. One \D meson is reconstructed using decay channels specific to the flavor of the decaying meson (see \cref{tab:sel:tagYields}). Usually multiple tag-signal candidate combinations are found, and we select the best combination of one tag and one signal candidate via the average beam-constrained mass $\mBC$ closest to the nominal \Dz mass. Finally, \mBC and \mKS are required to be within the axis boundaries of \cref{fig:sel:doubleTagmBCmKS}.

Tagged and untagged samples contain \num{1935} and \num{13209} candidates, respectively. The contributions to the tagged sample from individual tag channels are listed in \cref{tab:sel:tagYields}. The signal yields are determined using a two-dimensional fit to \mBC and \mKS.
\begin{table}
   \centering
   \caption{Yields of the tagged sample for various tag channels. Yields include background candidates.}
   \label{tab:sel:tagYields}
   \begin{tabular}{c|c}
		 \toprule
		 Flavor tag 	&  Signal yield\\
		 \midrule
		 \Km\kern -0.16em\pip 										              & \num{361} \\
		 \Km\kern -0.16em\pip\kern -0.16em\piz        			    & \num{702} \\
		 \Km\kern -0.16em\pip\kern -0.16em\ppz      				    & \num{178} \\
		 \Km\kern -0.16em\pip\kern -0.16em\pipi      				    & \num{517} \\
		 \Km\kern -0.16em\pip\kern -0.16em\pipi\kern -0.16em\piz& \num{114} \\
     \Km\kern -0.16em\pip\kern -0.16em\kern -0.16em$\eta$   & \num{63} \\
		 \midrule
		  Total & \num{1935} \\
		 \bottomrule
	  \end{tabular}
\end{table}

\subsection{Signal and background}
\label{sec:dataPrep:background}
\begin{figure}[tbp]
		\includegraphics[width=0.48\textwidth]{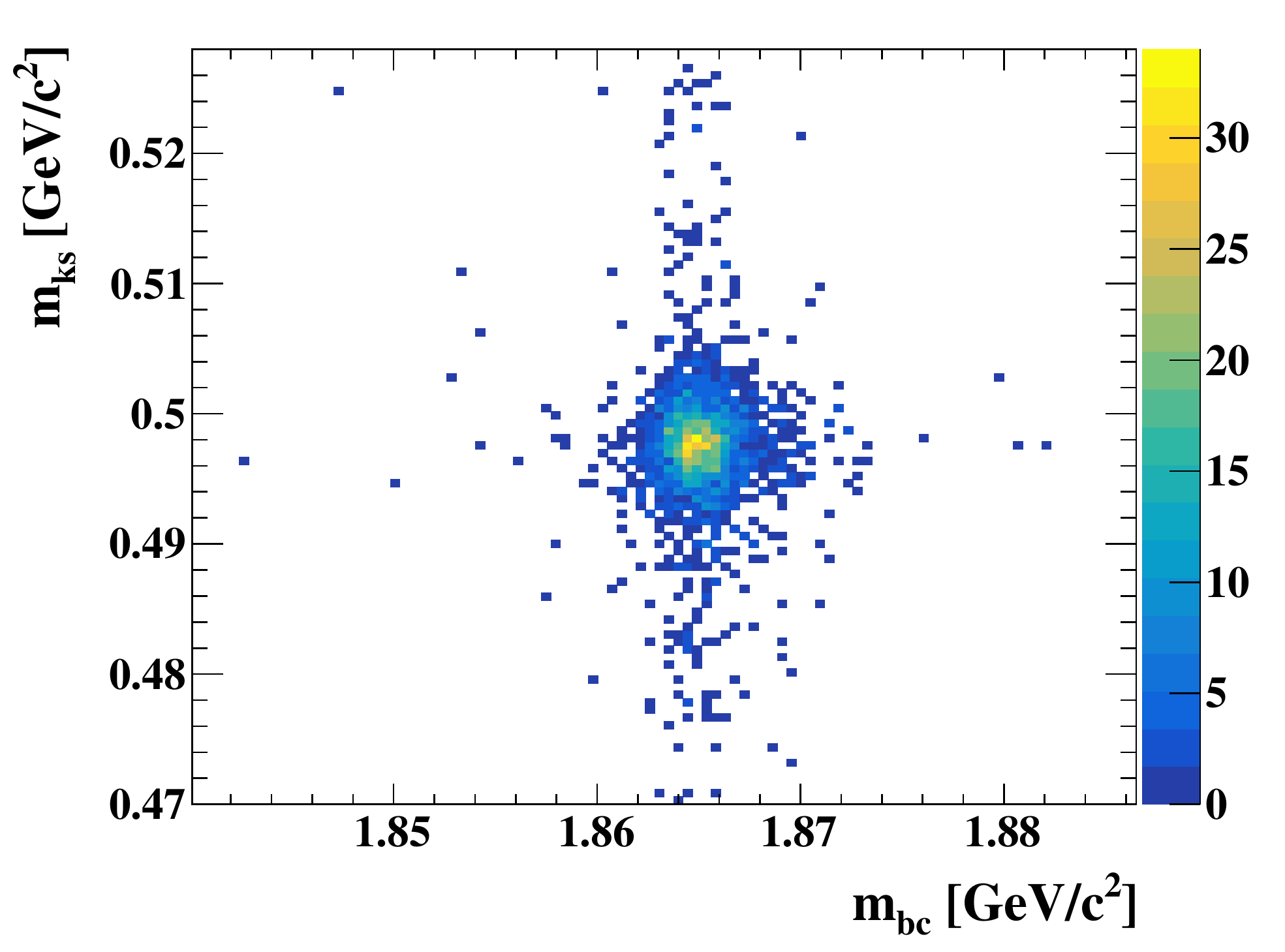}
		\caption{Distribution of \mKS versus \mBC of the untagged sample.}
	\label{fig:sel:doubleTagmBCmKS}
\end{figure}
The analysis requires accurate determination of the signal yields of tagged and untagged samples.
The background that passes our selection can be categorized according to its distribution in the plane of \mKS versus \mBC:
\begin{itemize}
  \item \textbf{Non-\KS background:} The final state of the signal decay is $\KpKm(\pipi)_{\KS}$. Events that do not contain the intermediate decay of a \KS show up as a peak in \mBC and a flat distribution in \mKS. 
	\item \textbf{Combinatorial background:} Events which do not contain the correct final state. These events come from \qqbar production and from misreconstructed \D decays. The \mBC distribution has a phase space component and a wide peak component. The \pipi pair can, but does not have to originate from a \KS decay. In case of an intermediate \KS decay the events show up as a band along \mKS. Otherwise they are broadly distributed.
\end{itemize}
The distribution of \mKS versus \mBC of the untagged sample is shown in \cref{fig:sel:doubleTagmBCmKS}. Since the signal peaks in the center of the \mBC versus \mKS plane, the signal and both background components can be distinguished in data by a fit procedure. 

We use an unbinned extended two-dimensional maximum likelihood fit to determine the signal yields.  Since the variables \mKS and \mBC are almost independent the two-dimensional probability density function (PDF) is constructed as a product of the one-dimensional PDFs. The signal components $S^{bc}(\mBC)$ and $S^{ks}(\mKS)$ are both modeled by a Crystal Ball function \cite{Gaiser:1982yw} with two-sided power law tails
\begin{align}
  S(x)=
  \begin{cases}
    (\frac{n_L}{|\alpha_L|})^{n_L}e^{-\frac{|\alpha_L|^2}{2}}(\frac{n_L}{|\alpha_L|}-|\alpha_L|-x)^{-n_L} \\
    e^{-\frac{x^2}{2}} \\
    (\frac{n_R}{|\alpha_R|})^{n_R}e^{-\frac{|\alpha_R|^2}{2}}(\frac{n_R}{|\alpha_R|}-|\alpha_R|+x)^{-n_R},
  \end{cases}
  \label{eqn:sel:crystalBall}
\end{align}
with $x=(m-\mu)/\sigma$. The three function components are defined for the lower tail $x < \alpha_L$, the central part $\alpha_L < x < \alpha_R$ and the higher tail $x > \alpha_R$. $S^{bc}$ and $S^{ks}$ have separate shape parameters $\mu, \sigma, n_R, n_L, \alpha_R$ and $\alpha_L$.

The combinatorial background model for \mBC consists of an \sftName{ARGUS} phase space shape $A(\mBC)$ \cite{Albrecht:1989ga} and a Gaussian $G_{c}(\mBC)$. The \mKS shape is described by a polynomial of first order $P^{(1)}$ and a peaking component modeled by the \mKS shape of the signal 
\begin{align}
  B_{c}^{bc}(\mBC) &= f^{bc}_{c} A(\mBC) + (1-f^{bc}_{c}) G_{c}(\mBC)\\
  B_{c}^{ks}(\mKS) &= f^{ks}_{c} P^{(1)}_{c}(\mKS) + (1-f^{ks}_{c}) S^{ks}(\mKS)\nonumber.
   \label{eqn:sel:pdfComb}
\end{align}

The non-\KS background model has the same shape as the signal in \mBC and a polynomial of first order $P^{(1)}$ in \mKS 
\begin{align}
  B_{k}^{bc}(\mBC) &= S^{bc}(\mBC) \\
  B_{k}^{ks}(\mKS) &=P^{(1)}_{k}(\mKS)\nonumber.
   \label{eqn:sel:pdfNonKs}
\end{align}
The untagged sample additionally contains non-\KS background candidates which were reconstructed from tracks of the 'tag' decay. We model these candidates by additional terms $A(\mBC)$ and $S^{ks}(\mKS)$ in $B_{k}^{bc}(\mBC)$ and $B_{k}^{ks}(\mKS)$, respectively. 

\begin{figure}[tbp]
  \centering
  \includegraphics[width=0.5\textwidth]{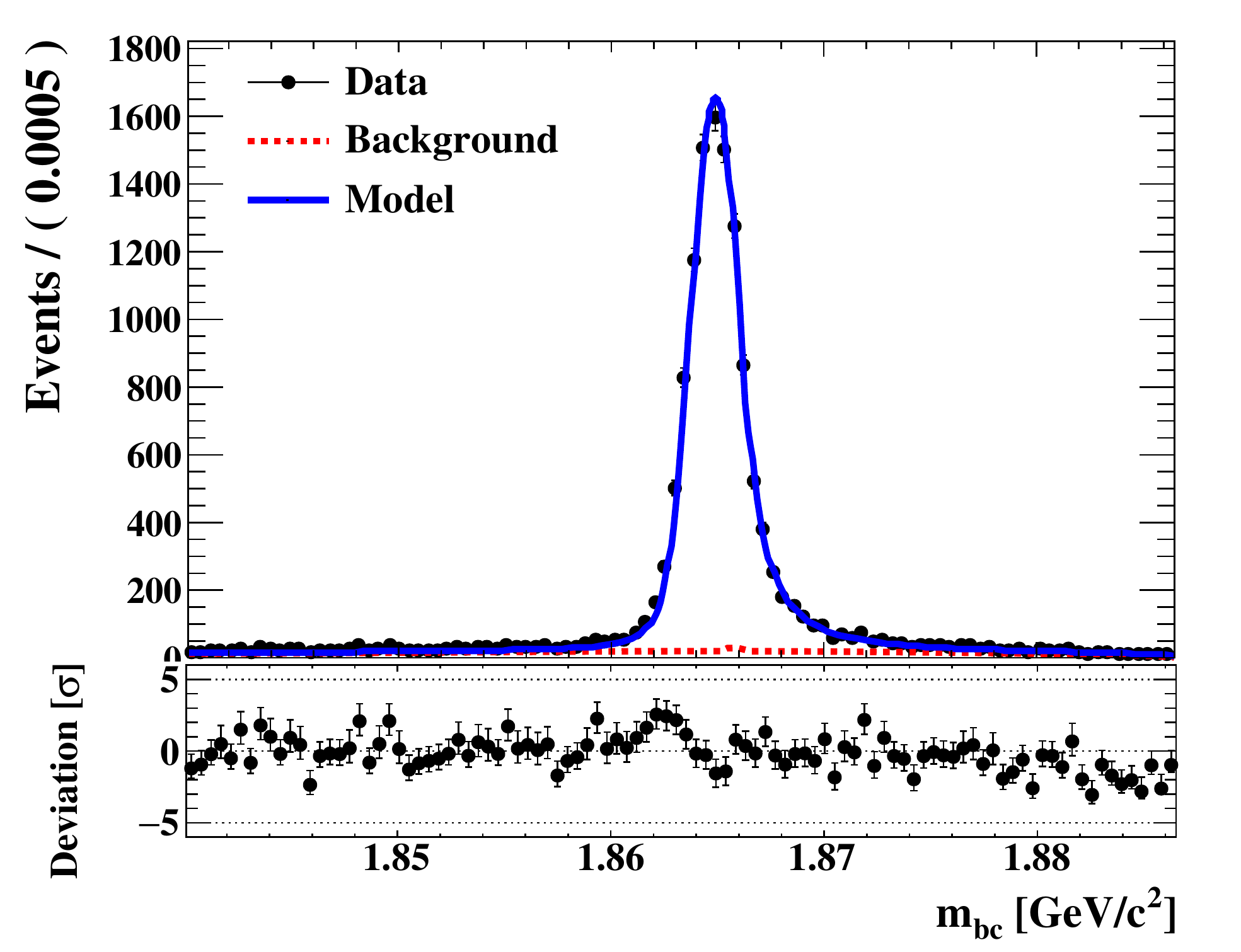}
  \includegraphics[width=0.5\textwidth]{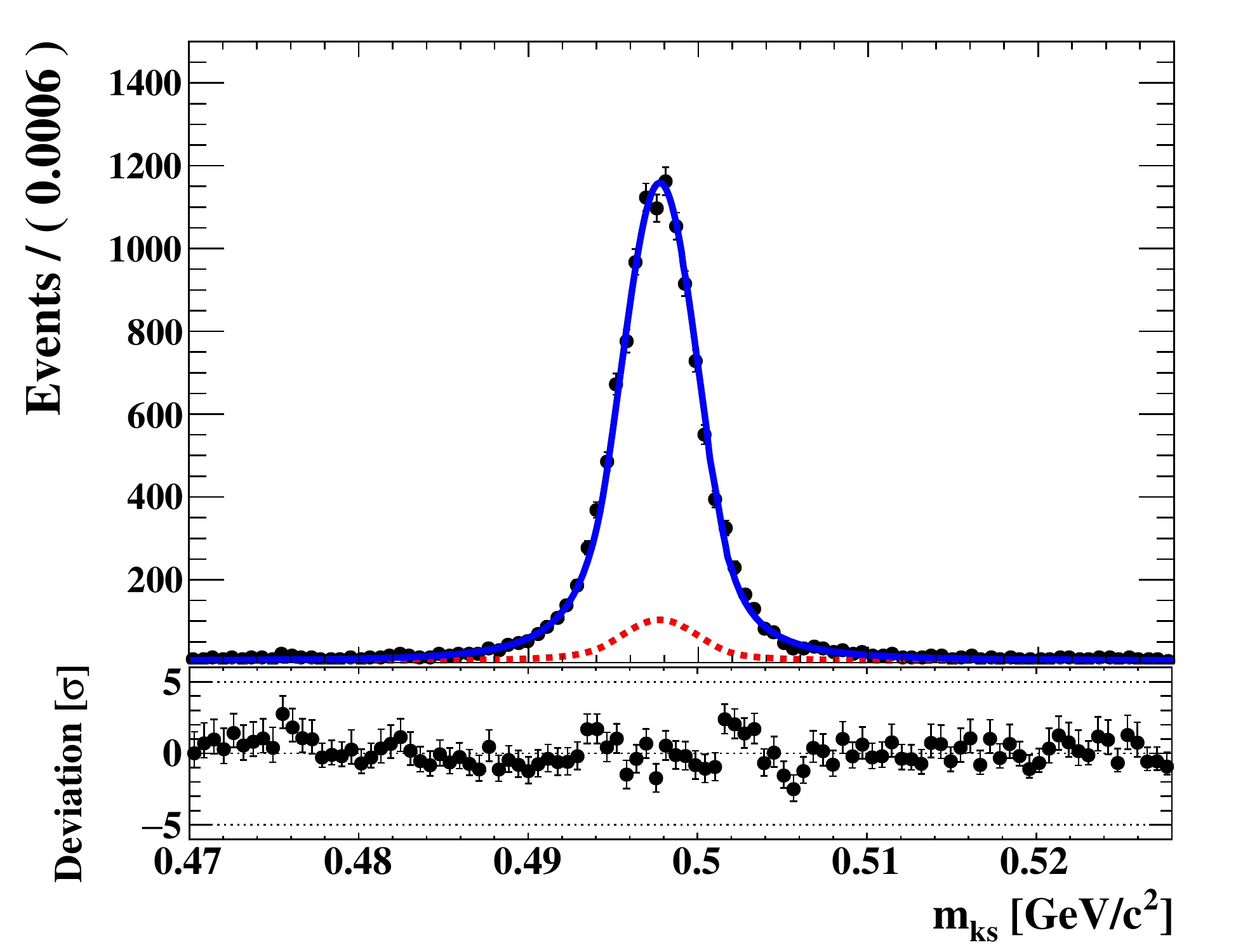}
  \caption{Projections of the untagged data sample and the fit model. Below the deviation between fit model and data sample is shown in units of its uncertainty.}
  \label{fig:bf:result}
\end{figure}

Shape parameters are determined using a simultaneous fit to MC samples representing signal and background components. The shape parameters, except the parameter $\sigma$ of the \KS signal peak, are fixed afterwards. The complete PDF is given by
\begin{align}
  F(\mBC,\mKS)=& N_{s}\ S^{bc}(\mBC) S^{ks}(\mKS) \nonumber\\
               &+ N_{c}\ B_{c}^{bc}(\mBC) B_{c}^{ks}(\mKS) \\
               &+ N_{k}\ B_{k}^{bc}(\mBC) B_{k}^{ks}(\mKS) \nonumber.
  \label{eqn:sel:totalPDF}
\end{align}
The yields $N_{s}$, $N_{c}$ and $N_{k}$ are determined by an extended maximum likelihood fit to the data sample. This procedure is used for the determination of the untagged signal yield for the branching fraction measurement. The tagged signal fraction for the Dalitz plot analysis is determined by a maximum likelihood fit with fixed sample yield. 
The fit to the untagged sample is shown in \cref{fig:bf:result} and yields \num{11660(118)} signal candidates.
For the tagged sample we require additionally to the description in \cref{sec:dataPrep:sel} that all candidates are within a box in the \mKS versus \mBC plane. We choose a box of $\pm\num{4}$ times the peak width $\sigma$ around the peak maximum in \mBC and \mKS, respectively. The signal yield determined by the fit within this box is \num{1856(45)} tagged signal candidates with a purity of \SI{96.37}{\percent}.

%% file: AmplitudeAnalysis.tex
\section{Dalitz plot analysis}
\label{sec:dalitz}
\begin{figure}[tbp]
   \centering
   \includegraphics[width=0.48\textwidth]{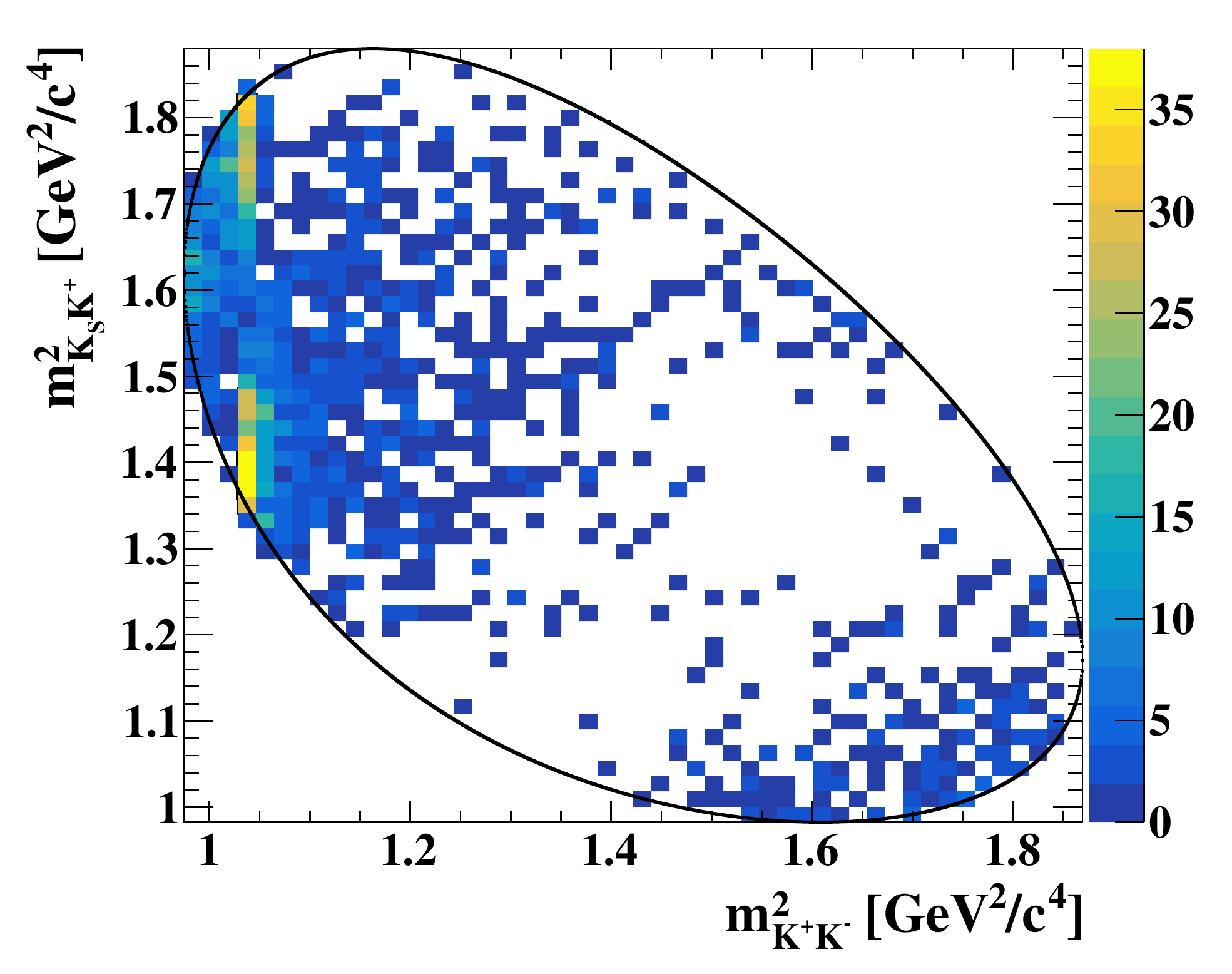}
   \caption{Dalitz plot of the decay \DKsKK from the tagged sample. The phase space boundary is indicated.}
   \label{fig:dalitz:data}
\end{figure}
The Dalitz plot of the decay \DKsKK after reconstruction and selection is shown in \cref{fig:dalitz:data}. Track momenta are updated according to the kinematic fit, described in \cref{sec:dataPrep:sel}. 
The distribution is dominated by the $\phi(1020)$ and the \KKbar $S$-wave which in turn is usually described by the charged and neutral $a_0(980)$ resonances. The $a_0(980)$ mass is below the \KKbar threshold and therefore only its high-mass tail is visible. From the distribution along the $\KS\Kp$ invariant mass the vector nature of the $\phi(1020)$ can be observed.

In the following, the free parameters of the Dalitz amplitude model are denoted with \FitPar and the Dalitz variables with \DPvar. The latter can be either two invariant masses (as used in \cref{fig:dalitz:data}) or an invariant mass and the corresponding helicity angle (see \cref{eqn:dalitz:helicityAngle}). 
The Dalitz plot analysis is performed using the \sftName{ComPWA} framework~\cite{Michel:2014nza, ComPWA:DKsKK}.

\subsection{Background}
\label{sec:dalitz:bkg}
\begin{figure}[tbp]
	\centering
	\includegraphics[width=0.5\textwidth]{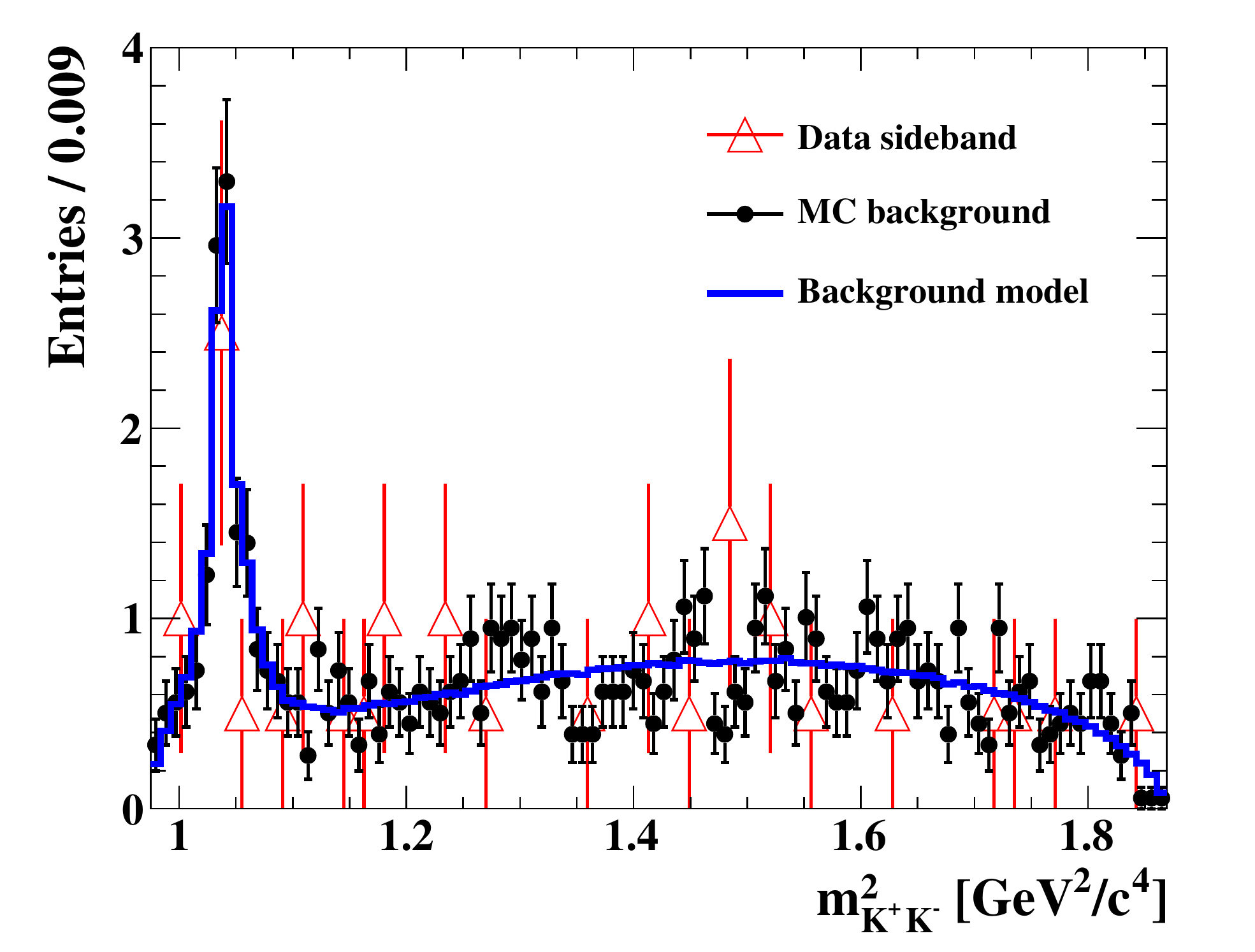}
	\caption{Projection of the MC background sample (full dotted points) and the background model (line) on \mKKsq, both scaled to the expected number of background candidates. The data sideband sample (open triangles) is scaled to approximate the MC background and is also re-binned due to low statistics.  Note that zero entries are not plotted.}
	\label{fig:dalitz:bkg:fit}
\end{figure}

The Dalitz plot of background candidates is studied using an MC background sample as well as data and MC sideband samples. Sideband sample events are required to lie outside a box region of $\pm 5$ times the peak width in \mBC and \mKS, and within $1.840 < \mBC < 1.8865$ GeV/c$^2$ and $0.470 < \mKS < 0.528$ GeV/c$^2$. The comparison of the MC samples with the data sample shows a good agreement (\cref{fig:dalitz:bkg:fit}), and the MC background sample is used in the following to fix the shape of a phenomenological background model. 
A peaking component originates from the decay \phiPiPi.  The $\phi(1020)$ is generated unpolarized; the effects of any spin alignment are expected to be negligible. We describe it by a Breit-Wigner model with mass and width parameters of the  $\phi(1020)$ and spin zero. An additional Breit-Wigner component with free parameters improves the fit quality close to the $\phi(1020)$ peak. Combinatorial background is described by a phase space component.
The background model shows good agreement with sideband data, as shown in \cref{fig:dalitz:bkg:fit}. We find $\chi^2/\text{ndf}= 116/99$ for the comparison of the model and sideband data.

\subsection{Quantum entangled \texorpdfstring{\DzDzb}{D0 anti-D0} decays}
\label{sec:dalitz:qc}
We analyze \Dz mesons produced in the reaction $\epem\to\DzDzb$ via a \psiprpr as intermediate state. In contrast to an isolated \Dz decay, this has implications for the decay rate since fundamental conservation laws hold for the combined \DzDzb decay amplitude and not just for the decay amplitude of one \Dz. We mention especially the conservation of charge-parity (\CP) which we assume to be strictly conserved in the \Dz system. We follow the phase convention 
\begin{align}
   \CP\ket{\Dz} = - \ket{\Dzb}.
\end{align}
The decay is mediated by the decay operator $\mathcal H$. In the following, the transition amplitude $\matrixel{j}{\mathcal H}{\Dz}$ of an isolated \Dz decay to the final state $j$ is denoted by ${\mathcal A}_j$. From \CP conservation it follows
\begin{align}
   {\mathcal A}_j = \matrixel{j}{\mathcal H}{\Dz} & = - \matrixel{\jmathBar}{\mathcal H}{\Dzb} = - \BAR{{\mathcal A}}_{\jmathBar} \nonumber \\
   \BAR{{\mathcal A}}_j = \matrixel{j}{\mathcal H}{\Dzb} & = - \matrixel{\jmathBar}{\mathcal H}{\Dz} = - {\mathcal A}_{\jmathBar}.
\end{align}
In the case that $j$ is a \CP eigenstate we have $j=\jmathBar$ and we include the \CP eigenvalue $\eta$ of the final state $j$: 
${\mathcal A}_j = - \eta \BAR{{\mathcal A}}_j$. 
We describe the amplitude ratio of \Dz and \Dzb to the same final state $j$ using its magnitude $r_j$ and phase $\delta_j$
\begin{align}
   \lambda_{j} = \frac{{\mathcal A}_j}{\BAR{\mathcal A}_j} = -r_j e^{-i\delta_j}.
   \label{eqn:dalitz:qc:lambda}
\end{align}
In general, the amplitudes ${\mathcal A}_j$ depend on the phase space position: ${\mathcal A}_j$ is constant only for two-body decays. 
We denote those final states with $r_j \leq 1$ by $j$ and their charge-conjugates with $r_{\jmathBar} > 1$ by $\jmathBar$.

The combined wave function of \DzDzb is anti-symmetric due to the negative parity of the \epem reaction. The matrix element of the decay of \Dz and \Dzb to the final states $i$ and $j$ at decay times $t_1$ and $t_2$, respectively, is given by
\begin{align}
	{\mathcal M}_{ij}(t_1,t_2) = & \frac{1}{\sqrt{2}} \bigg[ 
		 \matrixel{i}{\mathcal H}{\Dz(t_1)} \matrixel{j}{\mathcal H}{\Dzb(t_2)} \nonumber\\
		& - \matrixel{i}{\mathcal H}{\Dzb(t_1)} \matrixel{j}{\mathcal H}{\Dz(t_2)}
	\bigg].
	\label{eqn:dalitz:qc:transitionMatrix}
\end{align}
The \bes experiment does not give access to the \Dz decay time and therefore we are only interested in the time-integrated transition matrix element. The integration of \cref{eqn:dalitz:qc:transitionMatrix} over the \Dz decay time difference yields
\begin{align}
	\abs{{\mathcal M}_{ij}}^2 &= \int_{-\infty}^\infty\abs{{\mathcal M}_{ij}(\abs{t_2-t_1})}^2 \  d(\abs{t_2-t_1}) \nonumber \\
							&\approx \abs{\BAR{\mathcal A}_j {\mathcal A}_i - {\mathcal A}_j \BAR{\mathcal A}_i}^2.
	\label{eqn:dalitz:qc:timeIntAmplitude}
\end{align}
We choose the normalization such that $\abs{{\mathcal M}_{ij}}^2$ and $\abs{{\mathcal A}_i}^2$ are branching fractions when integrated over the phase space. The \Dz mixing parameters $x$ and $y$ are of \order~($10^{-3}$)~\cite[p.~691ff]{Tanabashi:2018oca} and thus are neglected in second order in the previous expression.

Using \cref{eqn:dalitz:qc:lambda} we can write \cref{eqn:dalitz:qc:timeIntAmplitude} as
\begin{align}
	\abs{{\mathcal M}_{ij}}^2 & \approx \abs{\BAR{\mathcal A}_j {\mathcal A}_i - {\mathcal A}_j \BAR{\mathcal A}_i}^2 \nonumber \\
		& = \abs{\BAR{\mathcal A}_i}^2\abs{\BAR{\mathcal A}_j\lambda_i - {\mathcal A}_j}^2.
\end{align}
For our case we get
\begin{align}
	\abs{{\mathcal M}_{\text{tag}, 3K}}^2	& = \abs{\BAR{\mathcal A}_{\text{tag}}}^2\abs{\BAR{\mathcal A}_{3K}\lambda_{\text{tag}} - {\mathcal A}_{3K}}^2.
	\label{eqn:dalitz:qc:DTMatrixEl}
\end{align}
The decay amplitudes of \Dz\to\KsKK and \Dzb\to\KsKK are connected via~\cite{Libby:2010nu}
\begin{align}
  \BAR{\mathcal A}_{3K}(\mKSKplus^2,&\mKSKminus^2) \nonumber\\
									& =	{\mathcal A}_{3K}(\mKSKminus^2,\mKSKplus^2)
\end{align}
if \CP is conserved. The amplitude of the flavor tag decay ${\mathcal A}_{\text{tag}}$  does not depend on the phase space position of the signal decay and is therefore constant in \cref{eqn:dalitz:qc:DTMatrixEl}. The ratio of \Dz to \Dzb amplitude of the tag decay is denoted by
\begin{align}
	\lambda_{\text{tag}} = -r_D e^{-i\delta_D}.
	\label{eqn:dalitz:lambdaTag}
\end{align}
We use experimental input for the magnitude and phase of $\lambda_{\text{tag}}$. Since these parameters have not yet been measured for each tag channel separately, we set them to common values for all tag channels.  As nominal value we choose the experimental average for the final state $\Km\pip$. Experimental results are summarized in \cref{tab:dalitz:HFAGMixing}.
\begin{table}
	\centering
  \caption{Hadronic parameters of tag channels. The mixing parameters and the hadronic parameters for the final state $\kaon\pi$ are obtained by a global fit~\cite{Amhis:2019ckw}. The measurement for $\kaon\pi\piz$ and $\Km\pip\pim\pip$ are from~\cite{Evans:2016tlp}. For the other tag channels no measurements exist today.}
	\label{tab:dalitz:HFAGMixing}
	\begin{tabular}{cc}
		\toprule
		Parameter & Value\\
		\midrule
    $x$                             & \SI{0.50(14)}{\percent}\\
    $y$                             & \SI{0.62(7)}{\percent}\\
    $\left(r_D^{\kaon\pi}\right)^2$ & \SI{0.344(2)}{\percent}\\
    $\delta_{K\pi}$                 & \SI{9.8(86)}{\degree}\\
    $r_D^{\kaon\pi\piz}$            & \SI{4.48(12)}{\percent}\\
    $\delta_{\kaon\pi\piz}$         & \SI{19(14)}{\degree}\\
    $r_D^{\particle{\kaon 3\pi}}$   & \SI{5.50(12)}{\percent}\\
    $\delta_{\kaon 3\pi}$           & \SI{-55(18)}{\degree}\\
		\bottomrule
	\end{tabular}
\end{table} 

A detailed derivation of the decay amplitude of quantum entangled \Dz mesons is given in~\cite{Tanabashi:2018oca,Asner:2005wf}.
To summarize, the result of an analysis of the \DKsKK Dalitz plot needs to be the amplitude model of an isolated \Dz decay in order to be comparable with other \Dz production reactions (\eg $\Dstar\to\Dz\pi$). 
Therefore, the effect of the production mechanism needs to be considered in the amplitude model.
For many \Dz final states this effect can be neglected, but it needs to be considered in this analysis since \KsKK is a self-conjugate final state.
The effect of the quantum entanglement on the measurement of the branching fraction is discussed in \cref{sec:bf:qc}.

\subsection{Resonance model}
\label{sec:dalitz:resonanceModel}
The signal decay amplitude ${\mathcal A}_{3K}$ is parameterized in the isobar model using the helicity formalism. The dynamic parts are described by a Breit-Wigner formula and, in the case of the \KKbar $S$-wave, by a \FLATTE description. We consider an intermediate resonance $R$ produced in the initial state $i$ and decaying to the final state $f$ with final state particles $a$ and $b$. The third (spectator) particle in the three-body decay is denoted by $c$. The resonance has the angular momentum $J$ and its parameterization depends on the center-of-mass energy squared $s$ of the final state particles $a$ and $b$.

The Breit-Wigner description suggested by the PDG~\cite{Tanabashi:2018oca} is
\begin{align}
	R^J(s) = -\frac{g_{D\to R}\ g_{R\to f}}{m_R^2-s+i\sqrt{s}\Gamma(s)},
	\label{eqn:th:resonances:relBreitWigner}
\end{align}
with the mass dependent width
\begin{align}
  \Gamma(s) = \Gamma_{R}\left(\frac{q(s)}{q(m_R^2)}\right)^{2J+1} \left(\frac{m_R}{\sqrt{s}}\right)\frac{F_{J}^2(z)}{F_{J}^2(z_{R})}.
	\label{eqn:th:resonances:gamma}
\end{align}

The center-of-mass daughter momentum $q(s)$ is imaginary below threshold. Therefore, we derive it from an analytic continuation of the phase-space factor:
  \begin{align}
    i\rho =
    \begin{cases}
      -\frac{\hat{\rho}}{\pi} \log{\abs{\frac{1+\hat{\rho}}{1-\hat{\rho}}}},          & s < 0\\
      -\frac{2\hat{\rho}}{\pi} \arctan{\frac{1}{\hat{\rho}}},                     & 0 < s <           s_{{\rm{th}}}\\
      -\frac{\hat{\rho}}{\pi} \log{\abs{\frac{1+\hat{\rho}}{1-\hat{\rho}}}}+i\hat{\rho},  &           s_{{\rm{th}}} < s\\
    \end{cases}.
    \label{eqn:th:dkskk:phasespace}
  \end{align}
with
\begin{align}
  \hat{\rho}(s)= \frac{1}{16\pi}\frac{2\sqrt{\abs{\hat{q}(s)}^2}}{\sqrt{s}},
  \label{eqn:th:kinematics:phasespacesimple}
\end{align}
and
\begin{align}
  \hat{q}^2(s) =  \frac{(s-(m_a+m_b)^2)(s-(m_a-m_b)^2)}{4s}.
  \label{eqn:th:kinematics:breakup}
\end{align}
\cref{eqn:th:kinematics:phasespacesimple} is input to \cref{eqn:th:dkskk:phasespace} 
and is in turn used to calculate $q(s)$ from the resulting $\rho$.
This is the parameterization suggested by the 
PDG~\cite[Section~48.2.3]{Tanabashi:2018oca}.

The coupling constants for the production and decay $g_{i}$ in \cref{eqn:th:resonances:relBreitWigner} are related to the partial width $\Gamma_i$ via
\begin{align}
	g_{R\to f} = \frac{1}{q^{J}(s_R) F_{J}(z_R)} \sqrt{\frac{m_R\Gamma_{R\to f}}{\rho(s)}}.
	\label{eqn:th:resonances:couplingToWidth}
\end{align}
This relation holds for narrow and isolated resonances.

The Breit-Wigner model assumes a point-like object. The effect of an extended resonance object is taken into account via the angular momentum barrier factors $F_{J}^2(z)$. The Blatt-Weisskopf barrier factors~\cite{blatt2012theoretical} are widely used: 
\begin{align}
	F_0^2(z) &= 1 \nonumber\\
	F_1^2(z) &= \frac{2z}{z+1} \\
	F_2^2(z) &= \frac{13z^2}{(z-3)^2+9z},\nonumber
\end{align}
where $z(s)=q(s)^2 R^2$. Experience shows that the influence of the resonance radius $R$ is rather small. We use $R = \SI{1.5}{\GeV^{-1}}$.

The \KKbar $S$-wave involves the charged and neutral $a_0(980)$. The $a_0(980)$ couples strongly to the channel \KKbar as well as to the channel $\eta\pi$. We describe both by a coupled channel formula, the so-called \FLATTE formula~\cite{Flatte:1976xu}. We use the parameterization from Ref.~\cite{Aubert:2005sm}:
\begin{align}
  R^J_{2ch}(s) = -\frac{g_{D\to R}\ g_{\KKbar}}{m_R^2-s+i(g_{\KKbar}^2\rho_{\KKbar}+g_{\eta\pi}^2\rho_{\eta\pi})}.
	\label{eqn:th:resonances:flatte}
\end{align}
The coupling constants are denoted by $g_i$ and the phase space factor $\rho_i$ is given in \cref{eqn:th:dkskk:phasespace}. The $a_0(980)^0$ couples also to the channel \KzKzb. Consequently, we add a third channel to the denominator of \cref{eqn:th:resonances:flatte} with the same coupling as the \KpKm channel. Other than this difference, the $a_0(980)^0$ and $a_0(980)^+$ use the same values for their coupling constants.  

The decay \DKsKK is a decay of a pseudo-scalar particle to three final state pseudo-scalar particles. The angular distribution of an intermediate resonance is therefore given by the Legendre polynomials $P_J(\cos\theta_R)$, which depend on the helicity angle
\begin{align}
   \cos\theta_{R} = -\frac{m^2_{ac}-m^2_a-m^2_c-2E_{R}^*E_{c}^*}{2q_{R}^*q_{c}^*}.
   \label{eqn:dalitz:helicityAngle}
\end{align}
Starred quantities are measured in the rest frame of the resonance.

The total decay amplitude ${\mathcal A}_j$ to a final state $j$ is the coherent sum over the individual resonances
\begin{align}
  {\mathcal A}_j(\DPvar, \FitPar)=\sum_{J}&\frac{\sqrt{2J\!+\!1}}{4\pi} \times \nonumber \\
								 &\sum_i c_i n_i R_i^J(s_i^2)P_J(\cos\theta_i).
\end{align}
The prefactor originates from the normalization of the Legendre polynomials. The Dalitz plot variables are denoted by \DPvar and consist of invariant mass $s_i^2$ and helicity angle $\theta_i$ for each subsystem. 

The magnitude and phase are given by the complex coefficient $c_i$, and all resonances are normalized with the factors
\begin{align}
	n_i^{-1} = \int \dd{\DPvar} R_i^J(\DPvar, \FitPar).
	\label{eqn:dalitz:resNorm}
\end{align}
We insert ${\mathcal A}(\DPvar, \FitPar)$ into \cref{eqn:dalitz:qc:DTMatrixEl} to obtain the decay amplitude including the effect of the quantum entanglement of \Dz and \Dzb. 
\subsection{Likelihood function}
\label{sec:dalitz:likelihood}
The probability function is given by
\begin{align}
  L(\DPvar,\FitPar) =& f\cdot\frac{ \abs{{\mathcal M}(\DPvar,\FitPar)}^2}{\int |{\mathcal M}(\DPvar',\FitPar)|^2\epsilon(\DPvar') \dd{\DPvar'}} \nonumber\\
  &+ (1-f)\cdot\abs{B(\DPvar)}^2.
  \label{eqn:dalitz:likelihoodFunc}
\end{align}
$B(\DPvar)$ denotes the Dalitz plot background model. The efficiency function is denoted by $\epsilon(\DPvar)$ and the signal purity by $f=(\num{96.37}\pm\num{0.43} ({\rm{stat.}}))~\si{\percent}$.
The normalization integrals are calculated with MC integration using a sample of phase space distributed candidates which have passed reconstruction and selection. In this way the efficiency correction is incorporated without the need to explicitly parameterize $\epsilon(\DPvar)$. The likelihood function is evaluated for each event, and the logarithm of its products can be written as
\begin{align}
	-\log{\lumi}(\FitPar) = -\sum_{\text{ev}}^{N} \log L(\DPvar_{\text{ev}},\FitPar),
	\label{eqn:dalitz:logLH}
\end{align}
where $N$ is the size of the data sample.
The interesting physics parameters are the fit fractions which are defined as
\begin{align}
	f_i = \frac{|c_i|^2 \int \dd{\DPvar} n_i^2 \abs{R_i(\DPvar, \FitPar)}^2}{\int \dd{\DPvar} \abs{\Amp(\DPvar, \FitPar)}^2},
	\label{eqn:dalitz:fitFractions}
\end{align}
where $c_i$ is the magnitude of resonance $R_i$. The integral $\int \dd{\DPvar} n_i^2 \abs{R_i(\DPvar)}^2$ is equal to one, due to our choice for the resonance normalization (\cref{eqn:dalitz:resNorm}).
A precise calculation of the statistical uncertainty of the fit fractions requires the propagation of the full covariance matrix through the integration which is achieved using an MC approach.

\subsection{Model selection}
\label{sec:dalitz:resSelection}
\begin{table}
	\centering
	\caption{Overview of resonances that could appear as intermediate states. Mass and width are the Breit-Wigner parameters. In case that these are channel dependent we quote the parameters of the \KKbar final state. The parameters of $f_0(980)$ and $a_0(980)$ are weighted averages of previous measurements of Refs.~\cite{Ambrosino:2006hb,GarciaMartin:2011jx,Ablikim:2004wn} and Refs.~\cite{Teige:1996fi, Abele:1998qd,Bugg:2008ig,Ambrosino:2009py,Athar:2006gq,Adams:2011sq}, respectively. The other values are from the PDG~\cite{Patrignani:2016xqp}.}
	\label{tab:dalitz:res:overview}
	\begin{adjustbox}{center}
		\resizebox{0.48\textwidth}{!}{
			\begin{tabular}{c|c|c|rl}
				\toprule
				Resonance 	& \IG(\JPC) 		& Mass~[\si{\mega\eVcSq}] 		& \multicolumn{2}{c}{Coupling} 		\\
				\midrule                            
				\multirow{2}{*}{$f_0(980)$} &\multirow{2}{*}{$0^+(0^{++})$}	& \multirow{2}{*}{\num{971(7)}}	& $g_{\KKbar} = $&\SI{3.54(5)}{\GeV} \\
				&				&					& $g_{\pi\pi} = $&\SI{1.5(1)}{\GeV}	\\
				$a_0(980)$ 							&$1^-(0^{++})$	&\numpme{994}{6}{4}& $g_{\eta\pi} = $&\SI{2.66(4)}{\GeV}\\
				$\phi(1020)$ 						&$0^-(1^{--})$	&\num{1019.461(19)} &$\Gamma = $&\SI{4.266(31)}{\MeV}\\
				$f_2(1270)$ 						&$0^+(2^{++})$	&\num{1275.5(8)}	&$\Gamma = $&\SIpme{185.9}{2.8}{2.1}{\MeV}\\ 
				$a_2(1320)$ 						&$1^-(2^{++})$	&\num{1318.1(7)}	&$\Gamma_{\KKbar} = $&\SI{109.8(24)}{\MeV}\\
				$f_0(1370)$~\cite{Vladimirsky:2006ky}&$0^+(0^{++})$	&\num{1440(6)} 		&$\Gamma_{\KKbar} = $&\SI{121(15)}{\MeV}\\
				$a_0(1450)$							&$1^-(0^{++})$	&\num{1474(19)}		&$\Gamma = $&\SI{256(13)}{\MeV}\\
				\bottomrule
			\end{tabular}
		} 
	\end{adjustbox}
\end{table}
The PDG~\cite{Tanabashi:2018oca} lists 12 resonances that could potentially contribute to the decay \DKsKK. Due to the limited size of the data sample we consider only resonance that were found to decay to \KKbar by previous experiments. An overview of known resonances is given in \cref{tab:dalitz:res:overview}. The choice of resonances that are included in the model is a common problem in amplitude analysis. Generally, increasing the complexity of the model improves the fit quality. In the usual approach, a minimum statistical significance or a minimum fit fraction (or both) is required for each resonance. Those requirements are somewhat arbitrary parameters and furthermore, the order in which resonances are added (or removed) from the model can lead to different sets of resonances. We apply a more abstract method for resonance selection.

Balancing a model between fit quality and model complexity is a common problem in the area of machine learning and in statistics in general. One approach to solve such a problem is the so-called Least Absolute Shrinkage and Selection Operator (LASSO) method~\cite{Tibshirani1996}. The basic idea is to penalize undesired behavior of the objective function. In the original approach the objective function is a least square model and the penalty function is the sum of the absolute values of the free parameters.
In the context of particle physics this approach is described in~\cite{Guegan:2015mea}. In our case the objective function is the logarithm of the likelihood, and the undesired behavior is a large sum over all fit fractions (which indicates strong interferences). Therefore, we use the sum of the square-root of fit fractions as penalty function. We modify \cref{eqn:dalitz:logLH}
\begin{align}
	\label{eqn:dalitz:resSelection:logLHpenalty}
	-\log{\lumi_{\lambda_P}}&(\FitPar) = -\log{\lumi}(\FitPar) \\
		 &+\lambda_P \sum_{i}\sqrt{\frac{\int\dd\DPvar\ N_i^2 R_i R_i^*}{\int\dd\DPvar\ \sum\limits_{m,n} N_m N_n R_m R_n^*}}. \nonumber
\end{align}
The square-root is used since it favors the suppression of small contributions, in contrast to, for example, the sum of the fit fractions which would favor solutions with equal values.

The parameter $\lambda_P$ regularizes the model complexity. A large value suppresses the sum of fit fractions and small values allow for larger interference terms. It is a nuisance parameter and we have to determine its optimal value. Again, this is a common problem in statistics and a possible solution is the use of so-called information criteria. Information criteria are mathematical formulations of the `principle of parsimony'~\cite{Schwarz1978}. This means in hypothesis testing that we prefer the model with fewer parameters over a more complicated model, given the same goodness-of-fit. 
The criteria suggested by~\cite{Guegan:2015mea} are the Akaike information criteria (AIC)~\cite{Akaike1974} and Bayesian information criteria (BIC)~\cite{Schwarz1978}. We choose a slightly modified version of the AIC which takes the size of the data sample into account:
\begin{align}
  {\rm{AIC}}^{\lambda_P}_c = -2\log{\lumi}_{\lambda_P}+2r_{\lambda_P}+\frac{2r(r+1)}{N-r-1}.
	\label{eqn:dalitz:resSelection:ic}
\end{align}
The number of events in data is denoted by $N$ and the coefficient $r$ is related to the complexity of the model. We follow the suggestion in Ref.~\cite{Guegan:2015mea} and use the number of resonances as parameter $r$. We consider only resonances with a fit fraction larger than a minimum value. A minimum value of \num{2e-3} gives a stable result. 

\begin{figure}[tbp]
	\includegraphics[width=0.5\textwidth]{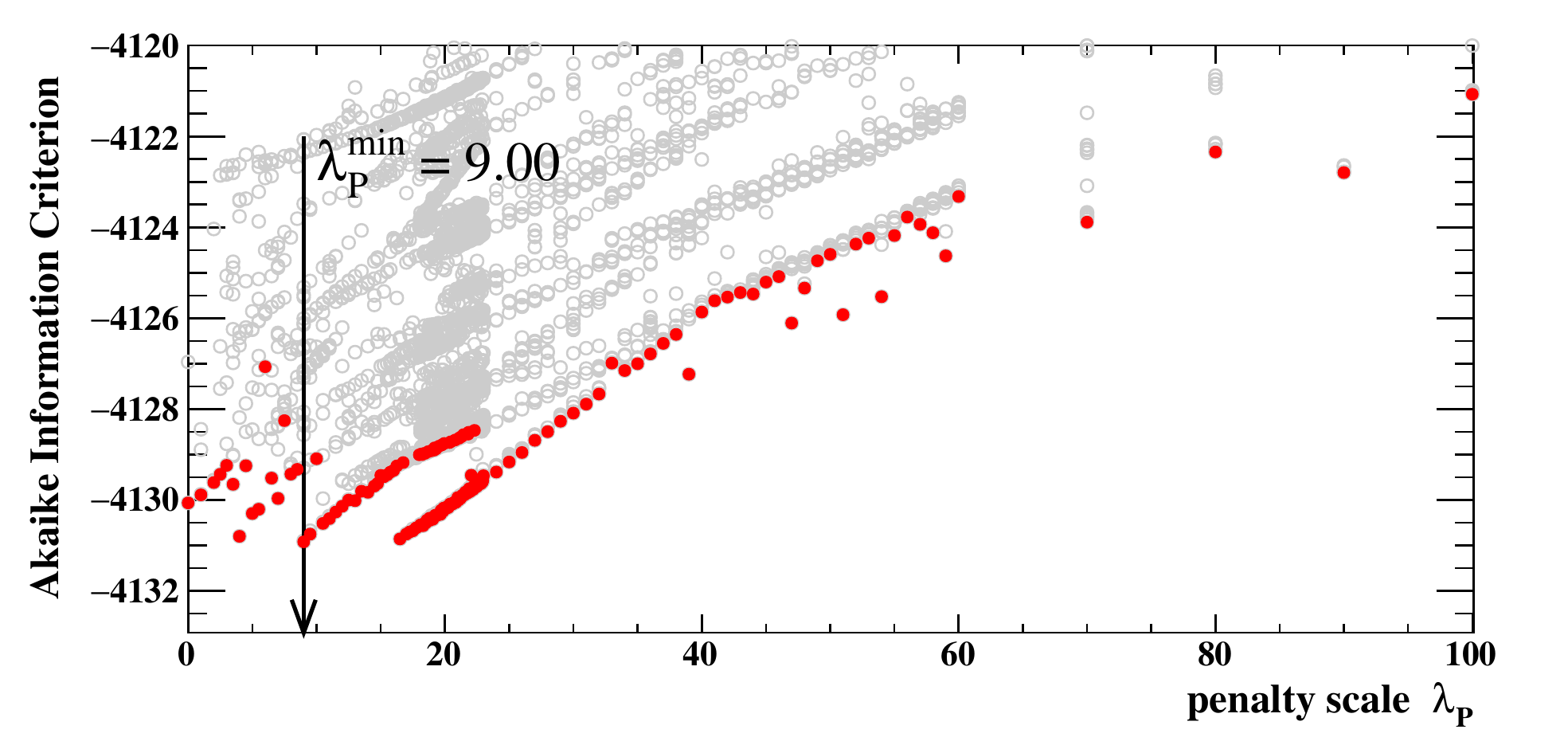}
	\caption{Distribution of $AIC^{\lambda_P}_c$ versus the penalty scale $\lambda_P$. A minimum is found at $\lambda_P^{min}=\num{9.0}$. For each value of $\lambda_P$ several fits with different starting parameters are performed (gray circles). Only the result with the lowest value (red dots) is considered.}
	\label{fig:dalitz:resSelection:penaltyScan}
\end{figure}

A scan for different values of $\lambda_P$ over a wide range is performed to map out the minima of $AIC^{\lambda_P}_c$. The scan is shown in \cref{fig:dalitz:resSelection:penaltyScan} and a minimum at $\lambda_P^{min}=\num{9.0}$ is found. The final set of resonances is stable versus small variations of $\lambda_P$. After a set of resonances is selected the penalty term is removed from the likelihood.

We obtain a model consisting of $a_0(980)^0$, $a_0(980)^+$, $\phi(1020)$, $a_2(1320)^+$, $a_2(1320)^-$ and $a_0(1450)^-$. The model contains the $a_2(1320)^-$ and $a_0(1450)^-$ which are expected to be doubly Cabibbo-suppressed with regard to their positively charged partner. We suspect that those contributions appear as an artefact of an imperfect model description in some phase space regions. We therefore decide to quote additionally a `basic' model of $a_0(980)^0$, $a_0(980)^+$ and $\phi(1020)$.

\subsection{Goodness-of-fit}
\label{sec:dalitz:gof}
The distribution of data events across the Dalitz plot is not uniform, and in a larger area of the phase space almost no events are observed (see \cref{fig:dalitz:data}). Therefore, we apply a goodness-of-fit test which is more suitable for this situation than the widely used $\chisq$ test. We choose a point-to-point dissimilarity method~\cite{Aslan2005} which provides an unbinned goodness-of-fit test. The test variable $\Phi$ is defined as 
\begin{align}
	\Phi = \frac{1}{2}\int &\dd\DPvar\int\dd\DPvar' \left[ \rho_m(\DPvar)-\rho_n(\DPvar)\right]\nonumber\\
	&\left[\rho_m(\DPvar')-\rho_n(\DPvar')\right] R(\abs{\DPvar - \DPvar'}).
	\label{eqn:dalitz:gof:phi1}
\end{align}
We use the Euclidean metric to calculate the distance between two points in phase space. For the general distance function $R(\abs{\DPvar - \DPvar'})$ we choose a Gaussian function with a width which is proportional to the amplitude value. 
The underlying (in general unknown) PDFs of two samples are denoted $\rho_m$ and $\rho_n$. Therefore, we estimate $\rho_m$ and $\rho_n$ by MC integration using samples from each PDF
\begin{align}
	\Phi &= \frac{1}{N(N+1)}\sum_{j>i}^N R(|\mathbf{n_i}-\mathbf{n_j}|)\nonumber\\
		 &- \frac{1}{NM}\sum_{j,i}^{M,N}R(|\mathbf{n_i}-\mathbf{m_i}|) \label{eqn:dalitz:gof:phi2} \\
		 &+ \frac{1}{M(M+1)}\sum_{j>i}^{M}R(|\mathbf{m_i}-\mathbf{m_j}|).\nonumber
\end{align}
Elements of both samples are denoted by $\mathbf{m_i}$ and $\mathbf{n_i}$ and the total sample size by $M$ and $N$, respectively. We identify one sample with our data sample, and the second one is generated using the final amplitude model. To calculate a probability that a certain model fits the data, the distribution of the test variable is needed. This can not be analytically derived and we simulate it using an MC approach; the result is given below.  

\subsection{Systematics}
\label{sec:dalitz:sys}
Systematic uncertainties on the Dalitz plot amplitude model arise from various sources: background description, amplitude model, inaccuracies of the MC simulation, external parameters and the fit procedure. For each source of uncertainty, we rerun the fit with a different configuration and add the deviations from the nominal amplitude model in quadrature.
An overview of the systematic uncertainties is given in \cref{tab:dalitz:sys:overview}.
\begin{table*}
  \centering
  \caption{Overview of uncertainties for the Dalitz plot amplitude model. We list the fit parameters and their statistical and systematic uncertainties. The fit parameters and fit fractions are corrected for their fitting biases and are denoted by `corrected value'. Systematic uncertainties are given in units of the statistical uncertainty of the parameter $\bar{\sigma}$ (we use the average value of the asymmetric uncertainties).}
  \label{tab:dalitz:sys:overview}
  \begin{adjustbox}{center}
	\resizebox{1.0\textwidth}{!}{
	  \begin{tabular}{c|c|c|ccc|ccc|ccc|ccc|ccc}
		\toprule
		\multirow{2}{*}{Parameter} & $g_{KK}$ & $a_0(980)^0$  &\multicolumn{3}{c|}{$a_0(980)^+$} &\multicolumn{3}{c|}{$\phi(1020)$} & \multicolumn{3}{c|}{$a_2(1320)^+$}& \multicolumn{3}{c|}{$a_2(1320)^-$}& \multicolumn{3}{c}{$a_0(1450)^-$} \\
                   &[\si{\GeV}]&FF~[\si{\percent}]&$\abs{c}$& $\phi~[\si{\radian}]$&FF~[\si{\percent}]&$\abs{c}$& $\phi~[\si{\radian}]$&FF~[\si{\percent}]&$\abs{c}$& $\phi~[\si{\radian}]$&FF~[\si{\percent}]&$\abs{c}$& $\phi~[\si{\radian}]$&FF~[\si{\percent}]&$\abs{c}$& $\phi~[\si{\radian}]$&FF~[\si{\percent}]\\
		\midrule
    Fit value					                    &3.80	&93	&0.59	&2.96	&33	&0.71	&1.67	&47	&0.13	&$-$2.97	&1.5	&0.10	&$-$0.14	&0.9	&0.21	&$-$0.23&4.1\\
    Corrected value 			                &3.77	&90	&0.64	&2.94	&34	&0.74	&1.67	&48	&0.12	&$-$2.92	&1.4	&0.09	&$-$0.06	&0.8	&0.16	&0.12	&2.2\\
		Mean stat.~uncertainty~$\bar{\sigma}$	&0.24	&10	&0.11	&0.17	&7	&0.06	&0.08	&2	&0.03	&0.23	  &0.6	&0.03	&0.23	  &0.4	&0.08	&0.58	&2.4\\
		Sys.~uncertainty 			                &0.35	&14	&0.09	&0.06	&6	&0.08	&0.19	&3	&0.01	&0.31	&0.3	&0.02	&0.28	&0.2	&0.04	&0.50	&1.9\\
		Total uncertainty 			              &0.42	&17	&0.14	&0.17	&9	&0.10	&0.21	&4	&0.03	&0.39	&0.7	&0.03	&0.36	&0.5	&0.10	&0.76	&3.1\\
	\midrule                                                                                                                                                                      
	\multicolumn{18}{c}{Systematic uncertainties in units of the mean statistical uncertainty $\bar{\sigma}$}\\
		\midrule                                                                                                                                                                      
    Background          &0.25	&0.71	&0.49	&0.18	&0.53	&0.49	&0.17	&0.29	&0.38	&0.34	&0.27	&0.11	&0.15	&0.17	&0.22	&0.53	&0.21\\
    Amplitude model   	&0.16	&0.27	&0.26	&0.17	&0.33	&0.18	&0.07	&0.09	&0.15	&0.04	&0.10	&0.04	&0.08	&0.12	&0.14	&0.28	&0.14\\
    Quantum correlation	&0.04	&1.21	&0.56	&0.24	&0.45	&1.14	&1.65	&1.91	&0.13	&0.32	&0.31	&0.19	&0.21	&0.12	&0.21	&0.52	&0.36\\
		External parameters &1.44	&0.49	&0.15	&0.17	&0.48	&0.36	&1.96	&0.27	&0.09	&1.27	&0.19	&0.56	&1.17	&0.59	&0.31	&0.34	&0.55\\
		Fitting procedure 	&0.07	&0.17	&0.20	&0.05	&0.13	&0.24	&0.05	&0.23	&0.12	&0.11	&0.08	&0.10	&0.17	&0.09	&0.29	&0.30	&0.41\\
		\midrule                                                                                                                                    
    Sys.~uncertainty    &1.46	&1.50	&0.79	&0.34	&0.86	&1.31	&2.56	&1.97	&0.43	&1.35	&0.46	&0.61	&1.21	&0.64	&0.53	&0.87	&0.80\\
		\bottomrule
	  \end{tabular}
	}
  \end{adjustbox}
\end{table*}

\subsubsection{Background}
\label{sec:dalitz:sys:bkg}
Uncertainties from the background treatment come from the background model as well as from the uncertainty on the signal purity.
The fit quality of the background model is good as illustrated in \cref{fig:dalitz:bkg:fit}, and we do not assign an uncertainty due to our choice of the model but we use different samples to determine the shape parameters. The nominal sample is the MC background sample and we additionally test MC and data sideband samples.
The difference of the fit result in comparison to the nominal model is taken as systematic uncertainty. Note that the contribution to the $\phi(1020)$ peak is different for the signal and sideband region. Thus, the systematic uncertainty is a conservative assumption.
The effect on the Dalitz plot analysis of the uncertainty on the signal purity is estimated by varying the signal purity by two times its statistical uncertainty to larger and smaller values. 

\subsubsection{Amplitude model}
\label{sec:dalitz:sys:ampModel}
A source of uncertainty of the amplitude model is the resonance radius that is used in the barrier factors. We vary it in steps of \SI{1}{\GeV^{-1}} from \SI{0}{\GeV^{-1}} to \SI{5}{\GeV^{-1}}. Our nominal value is \SI{1.5}{\GeV^{-1}}. 

The quantum entanglement of \DzDzb is included in the Dalitz amplitude model. We use external measurements of the magnitude and phase of $\lambda_{\text{tag}}$ (\cref{eqn:dalitz:lambdaTag}).
The experimental averages for $r_D$ and $\delta_D$ from the final state $\Km\pip$ are used as nominal values. The influence on the result is studied using the value for $r_D$ from $\Km\pip\piz$ and twice the $K^-\pi^+$ nominal value. The phase $\delta_D$ is set to zero, twice the $K^-\pi^+$ nominal value and to the measured value of $\Km\pip\pim\pip$.

\subsubsection{Monte-Carlo simulation}
\label{sec:dalitz:sys:dataMc}
The efficiency correction of the data sample is obtained from MC simulation. Differences between data and MC simulation in track reconstruction and particle identification can influence the result. Especially, regions with low momentum \Kpm tracks are prone to inaccuracies. We correct for these differences using momentum dependent correction factors obtained from hadronic \Dz decays. We test the influence of the tracking correction by rerunning the fit without correction. The influence is found to be negligible for the Dalitz plot analysis and thus no systematic uncertainty is assigned.

Another effect comes from different momentum resolutions in data and MC simulation. The $\phi(1020)$ has a width that is of the same order as the mass resolution. We study the influence of the mass resolution by rerunning the minimization with a free width parameter which approximates a resolution difference. The parameter changes from \SI{4.266}{\MeV} to \SI{5.2(3)}{\MeV}. We add the deviation from the nominal model to the systematic uncertainty. We keep the parameter fixed in the nominal fit.

\subsubsection{External parameters}
\label{sec:dalitz:sys:external}
External parameters are listed in \cref{tab:dalitz:res:overview,tab:dalitz:HFAGMixing}. We shift each parameter by its uncertainty to smaller and larger values and rerun the minimization. The deviation from the nominal model is taken as systematic uncertainty. The influence of the $a_0(980)$ coupling to $\eta\pi$ is estimated by rerunning the minimization with both couplings as free parameters. We obtain a value of $g_{\eta\pi}=\SI{2.54(16)}{\GeV}$.

\subsubsection{Fit procedure}
\label{sec:dalitz:sys:fitValidation}
We validate that the analysis routine is bias free and that the fit routine provides a correct estimate of the statistical uncertainty. We use our nominal fit result to generate a signal MC sample. This sample passes detector simulation and reconstruction as well as the event selection procedure. Then, we add the expected amount of background from MC simulation and rerun the minimization procedure. We calculate the difference between the parameters of the nominal model and the fit result in units of the statistical uncertainty of the parameter. The procedure is repeated with \num{200} statistically independent samples.

We find that the error estimate is correct but small biases for some parameters are present, especially for the parameters of the $a_0(1450)^-$. We correct each fit parameter for its bias and add half of the correction to the systematic uncertainty.

Furthermore, we check that no better minimum exists in the parameter space. We do so by rerunning the minimization with start values chosen randomly across the whole parameter space. From \num{200} fits, no fit with a valid minimum exhibits a smaller negative logarithm of the likelihood value than the nominal fit.

\subsection{Results}
\label{sec:dalitz:result}
\begin{figure}[tbp]
  \centering
  \includegraphics[width=0.5\textwidth]{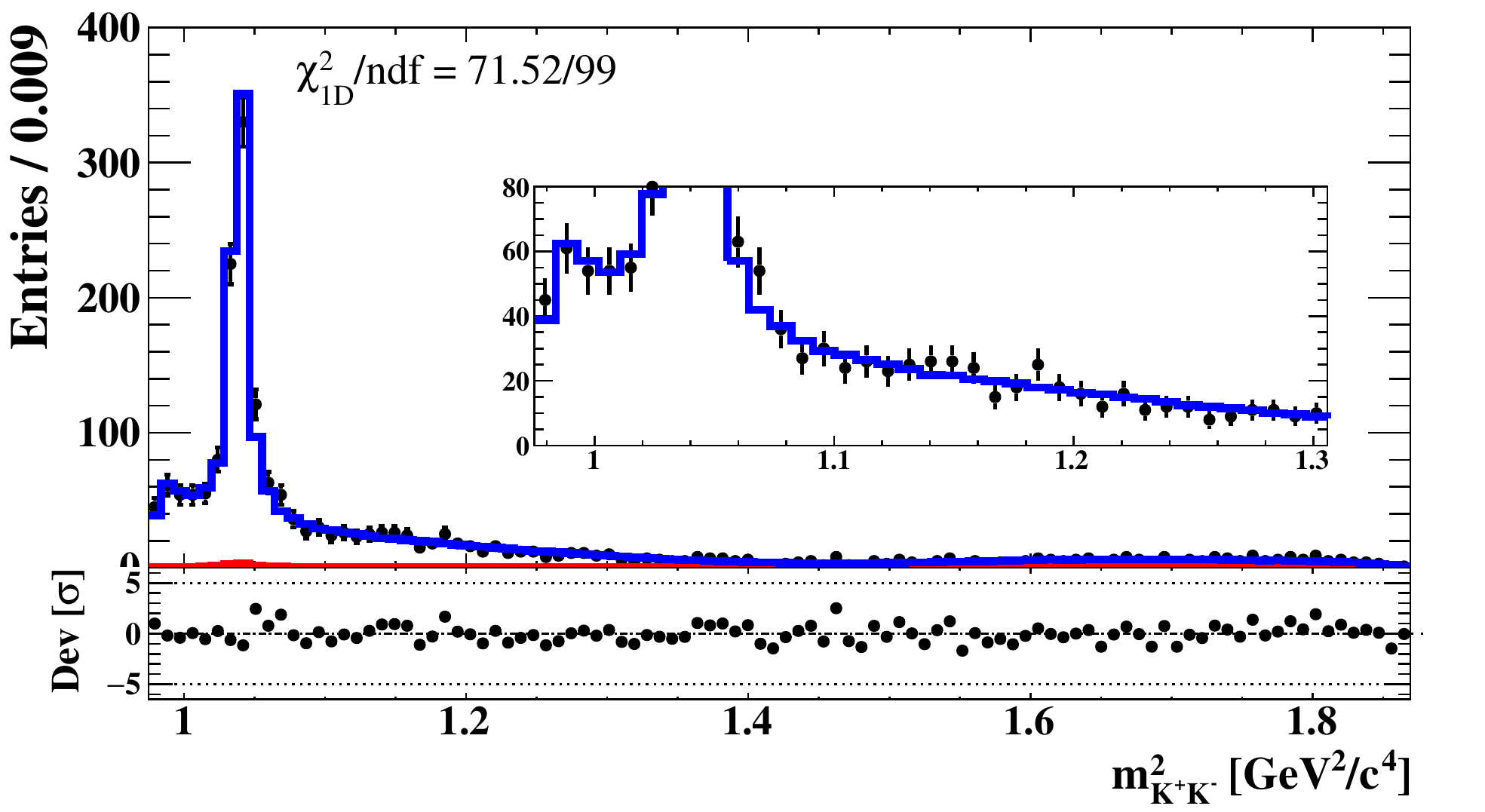}
  \includegraphics[width=0.5\textwidth]{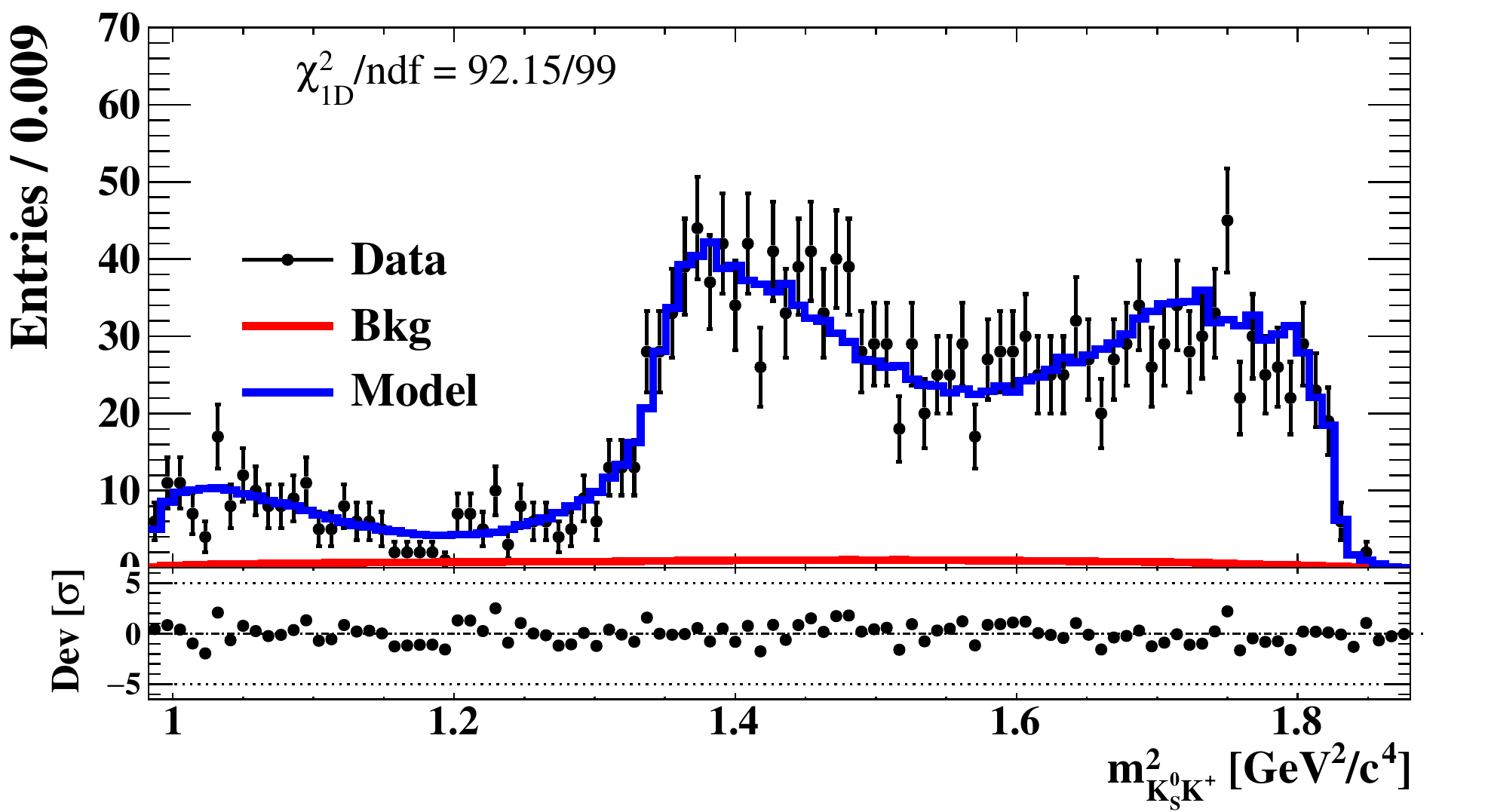}
  \includegraphics[width=0.5\textwidth]{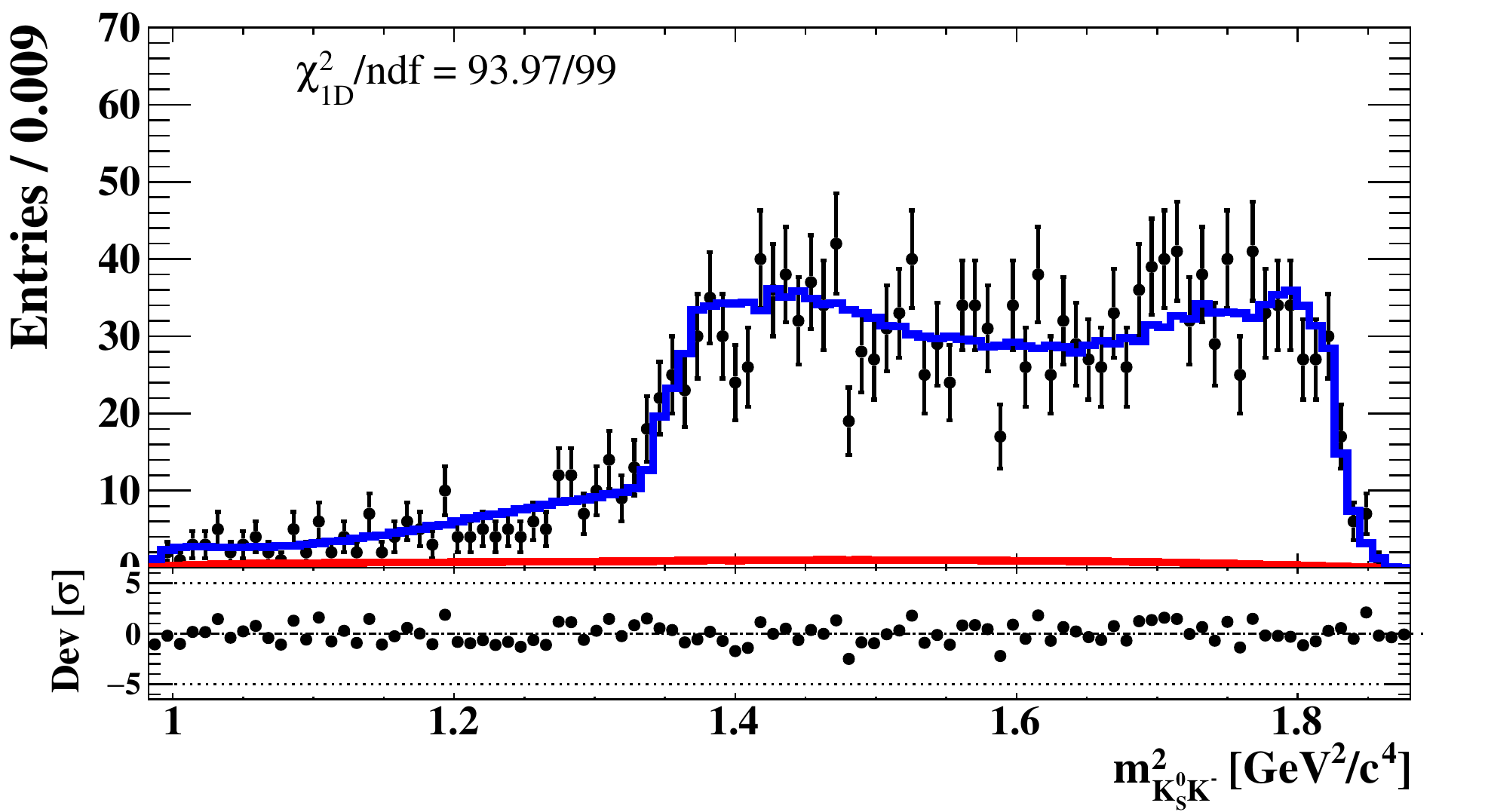}
  \caption{Dalitz plot projections of data sample (full dots) and amplitude model (blue line). Below each projection the deviation between model and data sample is shown in units of its uncertainty. The inset in the first plot shows a zoom on the \KKbar $S$-wave contribution.}
  \label{fig:dalitz:result}
\end{figure}

We find that the Dalitz plot is well described by a model with six resonances: $a_0(980)^0$, $a_0(980)^+$, $\phi(1020)$, $a_2(1320)^+$, $a_2(1320)^-$ and $a_0(1450)^-$. The Dalitz plot projections and the fit model are shown in \cref{fig:dalitz:result} and the fit parameters are listed in \cref{tab:dalitz:result}. The magnitude and phase of the $a_0(980)^0$ are fixed to \num{1} and \num{0}, respectively, as a reference. 
\begin{table*}[tbp]
  \centering
  \caption{Result from the Dalitz plot analysis. The first uncertainty is statistical followed by systematic uncertainty. The coupling constant $a_0(980)\to\KKbar$ is determined to be $g_{\KKbar}=\SIe{3.77}{0.24}[0.35]{\GeV}$. For the $a_2(1320)^+$, $a_2(1320)^-$ and $a_2(1320)^+$ the upper limits and the central values (CV) of the fit fractions are quoted as well as their combined significance.}
  \label{tab:dalitz:result}
  \begin{adjustbox}{center}
  \begin{tabular}{c|c|c|c|c}
    \toprule
    Final state &Magnitude 
                &Phase [\si{\radian}] 
                &Fit fraction [\si{\percent}]
                &Sign.[\si{\sigma}] \\
    \midrule
    $a_0(980)^0 \KS$	&\num{1}
                      &\num{0}
                      &\nume{90}{10}[17] 
                      &$>$\num{10}\hphantom{aaaaa}\\
    $a_0(980)^+ \Km$	&\numpme{0.64}{0.14}{0.08}[0.09] 
                      &\numpme{2.94}{0.19}{0.14}[0.06] 
                      &\nume{34}{7}[6]
                      &$>$\num{10}\hphantom{aaaaa}\\
    $\phi(1020) \KS$  &\numpme{0.74}{0.08}{0.04}[0.08] 
                      &\nume{1.67}{0.08}[0.19]  	      
                      &\nume{48}{2}[3]
                      &$>$\num{10}\hphantom{aaaaa}\\
    $a_2(1320)^+ \Km$	&\nume{0.12}{0.03}[0.01]         
                      &\numpme{-2.92}{0.21}{0.26}[0.31]
                      &$<$ \num{2.3} ($@$\SI{90}{\percent} C.L.), CV = \num{1.4}
                      &\multirow{3}{*}{\hphantom{$<$}$\begin{rcases} \num{3.9}\\ \num{3.5}\\ \num{3.5}\\ \end{rcases}\num{5.9}$}\\
    $a_2(1320)^- \Kp$	&\nume{0.09}{0.03}[0.02]         
                      &\nume{-0.06}{0.23}[0.28]        
                      &$<$ \num{1.6} ($@$\SI{90}{\percent} C.L.), CV = \num{0.8}
                      &\\
    $a_0(1450)^- \Kp$	&\numpme{0.16}{0.12}{0.05}[0.04] 
                      &\nume{0.12}{0.58}[0.50]         
                      &$<$ \num{13.2} ($@$\SI{90}{\percent} C.L.), CV = \num{2.2}
                      &\\
                      \midrule
    Total 				    &
                      &
                      &\nume{176}{20}
                      &\\
                      \bottomrule
  \end{tabular}
  \end{adjustbox}
\end{table*}

The projections of the model and the data sample show an excellent fit quality. The probability of the goodness-of-fit test of the model is \SI{71(3)}{\percent}. The fit fractions for the interference terms are listed in \cref{tab:dalitz:interference}. The largest interference is a destructive interference between the neutral and charged $a_0(980)$. The total fit fraction of the interference terms sums up to \SI{106}{\percent}.
\begin{table}
  \centering
  \caption{Fractions of interference terms in percent of the nominal amplitude model. Values are given without uncertainties and systematic corrections. Interference terms between \Dz and \Dzb amplitudes are omitted.}
  \label{tab:dalitz:interference}
  \sisetup{round-mode=places,round-precision=2,table-omit-exponent, fixed-exponent = -2} 
  \begin{adjustbox}{center}
    \resizebox{0.5\textwidth}{!}{
      \begin{tabular}{c|SSSSS}
        \toprule
      &  {$a_0(980)^+$} & {$\phi(1020)$} &  {$a_2(1320)^+$} & {$a_2(1320)^-$} & {$a_0(1450)^-$}\\
      \midrule
        $a_0(980)^0$  &-0.7483254442  &0.001202828387 &-0.01403488354 &0.0056438379 &-0.1405050955 \\ 
        $a_0(980)^+$  &&  -0.02931586468 &0.0004250921873 & -0.006248431554 &   0.08233559952 \\    
        $\phi(1020)$  &&&  -0.005905181425 & 0.0009543576763 &   0.01158218492 \\                     
        $a_2(1320)^+$ &&&& -0.006072473928 & -0.001087128602 \\                                      
        $a_2(1320)^-$ &&&&& -0.0002438285805 \\                                                     
        \bottomrule
      \end{tabular}
    }
  \end{adjustbox}
\end{table}

Furthermore, we study a model with a reduced set of resonances that only includes $a_0(980)^0$, $a_0(980)^+$, and $\phi(1020)$. The results are listed in \cref{tab:dalitz:result-basic}. The probability of the goodness-of-fit test of the reduced model is \SI{68(3)}{\percent}. The nominal amplitude model is used below in the branching fraction measurement to obtain the signal efficiency.

\begin{table}[tbp]
  \centering
  \caption{Result from the Dalitz plot analysis using a model with resonant contributions from $a_0(980)^0$, $a_0(980)^+$ and $\phi(1020)$. The first uncertainty is statistical followed by systematic uncertainty. The coupling constant $a_0(980)\to\KKbar$ is determined to be $g_{\KKbar}=\SIe*{3.47}{0.20}[0.38]{\GeV}$. }
  \label{tab:dalitz:result-basic}
  \begin{adjustbox}{center}
    \resizebox{0.5\textwidth}{!}{
      \begin{tabular}{c|c|c|c}
        \toprule
        Final state &Magnitude
                    &Phase [\si{\radian}]
                    &Fit fraction [\si{\percent}] \\
        \midrule
        $a_0(980)^0 \KS$	&\num{1}  					     
                          &\num{0}  						               
                          &\nume{79}{5}[7] \\
        $a_0(980)^+ \Km$	&\nume{0.67}{0.04}[0.07] 
                          &\numpme{-2.99}{0.06}{0.09}[0.19]   
                          &\nume{36}{3}[6] \\
        $\phi(1020) \KS$	&\nume{0.78}{0.03}[0.04] 
                          &\nume{1.84}{0.07}[0.20]  	         
                          &\nume{48}{1}[3] \\
        \midrule
        Total 				    &                        
                          &
                          &\nume{163}{11}\\
        \bottomrule
      \end{tabular}
    }
  \end{adjustbox}
\end{table}

%% file: BF.tex
\section{Branching fraction measurement}
\label{sec:bf}
The branching fraction of \DKsKK is measured using the untagged sample. The branching fraction is given by
\begin{align}
	\BR_{3K} = \frac{N^{3K}}{2 N_{\DzDzb}\cdot f_{QC}\cdot\epsilon_{3K}\cdot\BR_{\KS\to\pip\pim}}.
	\label{eqn:bf:BF}
\end{align}
Here, the signal yield is denoted by $N^{3K}$, which is corrected for the efficiency of reconstruction and selection $\epsilon_{3K}$. We correct for the branching fraction of the \KS reconstruction mode using $\BR(\KS\to\pip\pim)=\SI{69.20(5)}{\percent}$~\cite{Patrignani:2016xqp}. Our branching fraction result is normalized to the number of \DzDzb decays in the data sample~\cite{Ablikim:2018tay}:
\begin{align}
  N_{\DzDzb} = \SIe{10597}{28}[98]{\timesten\tothe{3}}.
\end{align}
The branching fraction in an untagged \DzDzb sample is linked to the branching fraction of an isolated \Dz decay via the correction factor $f_{QC}$ which is derived in the following section.
\subsection{Quantum entanglement}
\label{sec:bf:qc}
We consider a pair of \Dz mesons, of which one meson decays to the signal final state and the other to an arbitrary final state. In the following, $i$ and $j$ are different final states. The branching fraction is given by
\begin{align}
	\BR_{jX}  = \abs{{\mathcal M}_{jX}}^2 & = \sum_i \left( \abs{{\mathcal M}_{ji}}^2 + \abs{{\mathcal M}_{j\overline{\imath}}}^2 \right) \nonumber \\
										  &= \sum_i \left( \BR_{ji} + \BR_{j\overline{\imath}} \right).
	\label{eqn:bf:matrixEl}
\end{align}
We sum over all possible final states of one \Dz meson. As mentioned before we use a normalization in which the phase space integral over the norm of an amplitude corresponds to a branching fraction. Using \cref{eqn:dalitz:qc:lambda} we find that
\begin{align}
	\BR_{jX} &= \sum_i \BAR{\BR}_j \BAR{\BR}_i \bigg[ 1+\expval{r_i}+\expval{r_j}+\expval{r_i}\expval{r_j} \nonumber\\
	&\qquad -\expval{2\sqrt{r_i}\cos\delta_i}\expval{2\sqrt{r_j}\cos\delta_j}\bigg].
	\label{eqn:bf:bfX1}
\end{align}
Here, $\expval{\cdot}$ denotes phase space averaged values. 
The branching fractions of isolated \Dz decay sum up to one
\begin{align}
	 \sum_i \left(\BAR{\BR}_i + \BR_{i} \right) = \sum_i \left(\BAR{\BR}_i + \BAR{\BR}_i \expval{r_i} \right) = 1
	\label{eqn:bf:sumOne}
\end{align}
and the mixing parameter $y$ can be expressed as~\cite{Asner:2005wf}
\begin{align}
	y=2\sum_i \BAR{\BR}_i \expval{\sqrt{r_i}\cos\delta_i}.
	\label{eqn:bf:mixingY}
\end{align}
Thus, \cref{eqn:bf:bfX1} gives
\begin{align}
	\BR_{jX} = 	\BAR{\BR}_j \left[ 1 + \expval{r_j} - \expval{2\sqrt{r_j} \cos\delta_j} y \right].
	\label{eqn:bf:bfX2}
\end{align}
The correction factor that links the branching fraction of a quantum entangled \DzDzb pair $\BR_{jX}$ to the branching fraction of an isolated \Dz decay $\BR_j$ is then given by
\begin{align}
	2f_{QC} &= 1 + \expval{r_j} - y \expval{2\sqrt{r_j}\cos\delta_j}.
	\label{eqn:bf:fQCValue}
\end{align}
 The quantities $r_j$ and $\delta_j$ depend on the phase space position, and we use the phase space averaged values for the calculation of $f_{QC}$. From the Dalitz amplitude model a value of $f_{QC} = \num{1.035(15)}$ is obtained for the final state $j=\KsKK$. The statistical uncertainty of the Dalitz amplitude model is propagated to $f_{QC}$ via an MC approach. The limited statistics of the tagged sample cause a rather large uncertainty on $f_{QC}$ of \SI{1.45}{\percent}.
\subsection{Systematic uncertainties}
\label{sec:bf:systematics}
Systematic uncertainties on the branching fraction measurement arise from several sources. An overview is given in \cref{tab:bf:sys:overview}.

Deviations between data and MC simulation can lead to different resolutions in specific variables, thus leading to different efficiencies for the selection criteria. 
Most selection variables are already included in the uncertainty on track reconstruction  (see below).  For the remaining requirement on the \chisq of the \Dz vertex fit we find an uncertainty of \SI{0.8}{\percent}.
Furthermore, we see a small difference in the \KS mass resolution and therefore the \KS width is a free parameter in the fit. The \Dz mass resolution is consistent between data and MC simulation.

The branching fraction measurement requires the total efficiency for reconstruction and selection which is sensitive to the substructure of the decay. We use the Dalitz plot model to generate signal events which we use for efficiency determination. Since the fit quality of the Dalitz model is excellent we do not assign an additional uncertainty. 

The signal yield is determined using models for signal and background. The model shape is determined using MC simulation and discrepancies between data and simulation can therefore lead to a bias in the yield determination. 
We use the covariance matrix of the fit and a multi-dimensional Gaussian to generate sets of shape parameters and recalculate signal and background yields using these sets of parameters. We find that the systematic uncertainty is less than \SI{0.2}{\percent}. Furthermore, we check that the fit reproduces the correct values.

The systematic uncertainty of the \KS reconstruction efficiency is studied using $\jpsi\to\Kstarpm\Kmp$ and $\jpsi\to\phi\KS\Kmp\pipm$ control samples~\cite{Ablikim:2015qgt}. We assign an uncertainty of \SI{1.2}{\percent} for it.

The efficiency for charged track reconstruction and particle identification is studied using hadronic \DDbar decays\cite{Ablikim:2015ixa}. We assign \SI{1}{\percent} uncertainty per charged kaon track.

The systematic uncertainty, and also the total uncertainty of the measurement is dominated by the contributions due to track reconstruction and particle identification. In total the systematic uncertainty on the branching fraction measurement is \SI{3.5}{\percent}.

\subsection{Result}
\label{sec:bf:result}
\begin{table}
	\centering
	\caption{Overview of systematic uncertainties.}
	\label{tab:bf:sys:overview}
	\begin{tabular}{cc|c}
		\toprule
		\multicolumn{3}{c}{Systematic uncertainties~[\si{\percent}]}\\
		\midrule
		\multicolumn{2}{c|}{Quantum entanglement} 		&\num{1.45}\\
		\multicolumn{2}{c|}{Selection}					&\num{0.80}\\
		\multicolumn{2}{c|}{Signal/background model} 	&\num{0.20}\\
		\midrule
		\multirow{4}{*}{\rotatebox{90}{Efficiency}} 
						&\KS reconstruction 		&\num{1.20}\\
						&\Kpm tracking 			&\num{2.00}\\
						&\Kpm particle identification	&\num{2.00}\\
						&MC statistics 				&\num{0.22}\\
		\midrule
		\multirow{2}{*}{\rotatebox{90}{Ext.}} 
            & Number of \DzDzb decays &\num{1.00}\\
						&\BR(\KS\to\pip\pim)		&\num{0.07}\\
		\midrule
		&Total 			&\num{3.67}\\
		\bottomrule
	\end{tabular}
\end{table}
The signal yield $N^{3K}$ is determined by a two-dimensional fitting procedure. The projections to \mBC and \mKS of the data sample and the fit model are shown in \cref{fig:bf:result}.  We obtain a signal yield of \num{11660(118)} events. Using the inclusive MC sample we find an efficiency for reconstruction and selection of $\epsilon_{3K}=\SI{17.04(4)}{\percent}$ where the uncertainty is due to limited MC statistics. 

According to \cref{eqn:bf:BF} the branching fraction of \mbox{\DKsKK} is
\begin{flalign}
	\BR(\DKsKK) & = \\
			 & \hspace{-2.5cm}\SIe*{4.51}{0.05}[0.16]{\timesten\tothe{-3}}\nonumber.
	\label{eqn:bf:result}
\end{flalign}
The relative statistical and systematic uncertainties are \SI{1.0}{\percent} and \SI{3.67}{\percent}, respectively. The total uncertainty is \SI{3.81}{\percent}. 

%% file: Conclusion.tex
\section{Conclusion}
\label{sec:conclusion}
In summary, we investigate the decay \DKsKK using \SI{2.93}{\invfb} of \epem collisions collected at $\sqrt{s}=\SI{3.773}{\GeV}$ recorded with the \bes experiment. We analyse the Dalitz plot and measure its branching fraction.

The \KsKK Dalitz plot is described using an isobar amplitude model. We select the optimal set of resonances using a `penalty term' method and find that the Dalitz plot is well described using an amplitude model with six resonances of $a_0(980)^0$, $a_0(980)^+$, $\phi(1020)$, $a_2(1320)^+$, $a_2(1320)^-$ and $a_0(1450)^-$. The largest contribution to the total intensity comes from the $a_0(980)^0$ that, together with its charged partner, describes the \KKbar threshold. Both resonances show a strong interference which leads to a sum of fit fractions of the Dalitz amplitude model of \SI{176(20)}{\percent}. 

The $f_0(980)$ could appear as an intermediate resonance, but our strategy for resonance selection does not favor a model that includes the $f_0(980)$. With respect to the nominal model its significance is \SI{1.02}{\sigma}. The $a_0(980)$ couples strongly to the channel \KKbar as well as to the channel $\pi\eta$. We measure its coupling to \KKbar to be $g_{\KKbar}=\SIe{3.77}{0.24}[0.35]{\GeV}$; within the uncertainties this is in agreement with previous measurements. For the Dalitz plot analysis, both \Dz mesons in each event are reconstructed. Therefore the sample size is limited and statistical and systematic uncertainties are of the same order. The result is influenced by the quantum entanglement of \Dz and \Dzb with respect to the measurements of isolated \Dz decays. We include this effect in our amplitude models in order to quote parameters of an isolated \Dz decay. The magnitude and phase of the ratio of \Dz and \Dzb amplitudes of the tag decays are necessary to describe this effect. Since those are not measured for all tag channels we use the parameters of \Dz\to\Kp\pim for all channels. The effect of this substitution on the final result is included in the systematic uncertainties.

The model includes the $a_2(1320)^-$ and the $a_0(1450)^-$ which are expected to be doubly Cabibbo-suppressed and, therefore, should have a significantly smaller fit fraction than their positively charged partners. The fact that we do not see this could be a hint that those contributions are artefacts of an imperfect model description in parts of the phase space. Those states and the $a_2(1320)^+$ have a combined statistical significance of \SI{5.9}{\sigma}. Each of their isospin partners $a_2(1320)^0$, $a_0(1450)^0$ and $a_0(1450)^+$ have a statistical significance of \SI{2.1}{\sigma} or less with respect to the nominal model and are not included by our method for resonance selection. Because of the small fit fractions of $a_2(1320)^+$, $a_2(1320)^-$ and $a_0(1450)^-$ we decide to report upper limits.
Due to these problems of the model, we additionally quote a model built from the `visible' resonant states $a_0(980)^0$, $a_0(980)^+$ and $\phi(1020)$. The result is given in \cref{tab:dalitz:result-basic}.
In comparison with the result from \babar~\cite{Aubert:2005sm} we use a different set of resonances which leads to stronger interference terms. 

We measure the branching fraction of the decay \DKsKK to be \SIe*{4.51}{0.05}[0.16]{\timesten\tothe{-3}}. This is the first absolute measurement.
We use the Dalitz amplitude model to accurately describe the signal decay in simulation and also to obtain the quantum entanglement correction factor. The measurement is in good agreement with previous measurements and we are able to reduce the uncertainty significantly. The measurement is systematically limited.

%% file: acknowledgements.tex
\section{Acknowledgments}
\label{sec:Acknowledgments}
The BESIII collaboration thanks the staff of BEPCII and the IHEP computing center for their strong support. This work is supported in part by National Key Basic Research Program of China under Contract No. 2015CB856700; National Natural Science Foundation of China (NSFC) under Contracts Nos. 11625523, 11635010, 11735014, 11822506, 11835012; the Chinese Academy of Sciences (CAS) Large-Scale Scientific Facility Program; Joint Large-Scale Scientific Facility Funds of the NSFC and CAS under Contracts Nos. U1532257, U1532258, U1732263, U1832207; CAS Key Research Program of Frontier Sciences under Contracts Nos. QYZDJ-SSW-SLH003, QYZDJ-SSW-SLH040; 100 Talents Program of CAS; INPAC and Shanghai Key Laboratory for Particle Physics and Cosmology; ERC under Contract No. 758462; German Research Foundation DFG under Contracts Nos. Collaborative Research Center CRC 1044, FOR 2359; Istituto Nazionale di Fisica Nucleare, Italy; Koninklijke Nederlandse Akademie van Wetenschappen (KNAW) under Contract No. 530-4CDP03; Ministry of Development of Turkey under Contract No. DPT2006K-120470; National Science and Technology fund; STFC (United Kingdom); The Knut and Alice Wallenberg Foundation (Sweden) under Contract No. 2016.0157; The Royal Society, UK under Contracts Nos. DH140054, DH160214; The Swedish Research Council; U. S. Department of Energy under Contracts Nos. DE-FG02-05ER41374, DE-SC-0010118, DE-SC-0012069; University of Groningen (RuG) and the Helmholtzzentrum fuer Schwerionenforschung GmbH (GSI), Darmstadt.